\documentclass[a4paper,10pt]{article}
\usepackage[utf8]{inputenc}
\usepackage{amssymb}									
\usepackage{listliketab}								
\usepackage{multirow}									
\usepackage{amsmath,amsthm}
\usepackage{graphicx}
\usepackage{float}
\usepackage{rotating} 
\usepackage{hyperref}
\hypersetup{colorlinks,%
citecolor=black,%
filecolor=black,%
linkcolor=black,%
urlcolor=black,%
}

\title{The Protein Family Classification in Protein Databases
via Entropy Measures\footnote{3rd version -- (June 30, 2017)}}
\author{R.P.~Mondaini, S.C.~de Albuquerque Neto}

\date{}

\begin{document}

\maketitle

\begin{abstract}
In the present work, we review the fundamental methods which have
been developed in the last few years for classifying into families
and clans the distribution of amino acids in protein databases.
This is done through functions of random variables, the Entropy
Measures of probabilities of the occurrence of the amino acids. An
intensive study of the Pfam databases is presented with restriction
to families which could be represented by rectangular arrays of
amino acids with \emph{m} rows (protein domains) and \emph{n}
columns (amino acids). This work is also an invitation to
scientific research groups worldwide to undertake
the statistical analysis with arrays of different numbers of rows
and columns, since we believe in the mathematical characterization
of the distribution of amino acids as a fundamental insight on the
determination of protein structure and evolution.
\end{abstract}

\section{Introduction and Motivation}

DESIDERATA: ``To translate the information contained on protein
databases in terms of random variables in order to model a dynamics
of folding and unfolding of proteins''.

The information on the planetary motion has been annotated on Astronomical
almanacs (Ephemerides) along centuries and can be derived and analyzed
by Classical Dynamics and Deterministic Chaos as well as confirmed and
corrected by General Relativity.  The information which is accumulated
on Biological almanacs (Protein databases) on the last decades, is
still waiting for its first description to be done by a successful
theory of folding and unfolding of proteins.  We think that this study
should be started from the evolution of protein families and its
association into Family clans as a necessary step of their development.

The first  fundamental idea to be developed here is that proteins do not
evolute independently. We introduce several arguments and we have
done many calculations to span the bridge over the facts about
protein evolution in order to emphasize the existence of a protein
family formation process (PFFP), a successful pattern recognition method
and a coarse-grained protein dynamics are driven by optimal control
theory \cite{mondaini1,mondaini2,mondaini3,mondaini4,mondaini5,mondaini6}.
Proteins, or their ``intelligent'' parts,
protein domains, evolute together as a family of protein domains. We then
realize that the exclusion of the evolution of an ``orphan'' protein is
guaranteed by the probabilistic approach to be introduced in the present
contribution. We think that the elucidation of the nature of intermediate
stages of the folding/unfolding dynamics, in order to circumvent the
Levinthal ``paradox'' \cite{levinthal,karplus} as well as the determination
of initial conditions, should be found from a detailed study of this PFFP
process. A byproduct of this approach is the possibility of testing the
hypothesis of agglutination of protein families into Clans by rigorous
statistical methods like ANOVA \cite{deGroot,mondaini4}.

We take many examples of Entropy Measures as the generalized
functions of random variables on our modelling. These are the
probabilities of occurrence of amino acids in rectangular arrays
which are the representatives of families of protein domains.
In section 2, the sample space which is adequate for this
statistical approach is described in detail. We start from the
definition of probabilities of occurrence and the restrictions
imposed on the number of feasible families by the structure of
this sample space. Section 3 introduces the set of Sharma-Mittal
Entropy Measures \cite{sharma,mondaini4} to be adopted as the
functions of probabilities of occurrence in the statistical
analysis to be developed. The Mutual Information measures
associated with the Sharma-Mittal set, as well as the normalized
Jaccard distance measures, are also introduced in this section.
In section 4, we present a naive sketch of assessing Protein
database, to set the stage for a more efficient approach of the
following sections. In section 5, we point out the inconvenience
of the Maple computing system for the statistical calculations to
be done, by displaying tables with all CPU and real times
necessary to perform all necessary calculations. We have also
provided in this section, some adaptation of our methods in order
to be used with the Perl computing system and we compare the new
times of calculation with those by using Maple at the beginning
of the section. We also include some comments on the use of Perl,
especially on its oddness to calculate with the input data given
in arrays and the way of circumventing this. However, we also stress
that despite the fact that joint probabilities and their powers could be
usually calculated, the output will come randomly distributed and the
CPU and real times will increase too much to favour
the calculation of the entropy measures. This is due to the
intrinsic ``Hash'' structure \cite{cormen} of the Perl computing system.
We then introduce a modified array structure in order to calculate with
Perl.

\section{The Sample Space for a Statistical Treatment}
We consider a rectangular array of \textbf{\emph{m}} rows (protein domains)
and \textbf{\emph{n}} columns (amino acids). These arrays are organized
from the protein database whose domains are classified into families and
clans by the professional expertise of senior biologists \cite{finn1,finn2}.

The random variable is the probability of occurrences of amino
acids, $p_j(a)$, $j=1,2,\ldots,n$, $a=A,C,D,\ldots,W,Y$ (one-letter
code for amino acids), to be given by
\begin{equation}
 p_j(a) \equiv \frac{n_j(a)}{m} \label{eq:eq1}
\end{equation}

\noindent where $n_j(a)$ is the number of occurrences of the amino acid
\textbf{\emph{a}} in the $j$-th column. Eq.(\ref{eq:eq1}) could be
also interpreted as the components of n vectors of 20 components
each
\begin{equation}
 \begin{pmatrix}
  p_1(A) \\ \relax
  \vdots \\ \relax
  p_1(Y)
 \end{pmatrix}
 \begin{pmatrix}
  p_2(A) \\ \relax
  \vdots \\ \relax
  p_2(Y)
 \end{pmatrix}
 \ldots
 \begin{pmatrix}
  p_n(A) \\ \relax
  \vdots \\ \relax
  p_n(Y)
 \end{pmatrix} \label{eq:eq2}
\end{equation}

\noindent and we have
\begin{equation}
 \sum_a n_j(a)=m\,, \,\, \forall j \, \Rightarrow \,
 \sum_a p_j(a)=1\,, \,\, \forall j \label{eq:eq3}
\end{equation}

Analogously, we could also introduce the joint probability
of occurrence of a pair of amino acids \textbf{\emph{a}}, \textbf{\emph{b}}
in columns \textbf{\emph{j}}, \textbf{\emph{k}}, respectively $P_{jk}(a,b)$
as the random variables. These are given by
\begin{equation}
 P_{jk}(a,b) = \frac{n_{jk}(a,b)}{m} \label{eq:eq4}
\end{equation}

\noindent where $n_{jk}(a,b)$ is the number of occurrences of the pair
of amino acids \textbf{\emph{a}}, \textbf{\emph{b}} in columns
\textbf{\emph{j}}, \textbf{\emph{k}}, respectively.

A convenient interpretation of these joint probabilities could be
the elements of $\frac{n(n-1)}{2}$, square matrices of $20 \times 20$
elements, to be written as
\begin{equation}
 P_{jk} =
 \begin{pmatrix}
  P_{jk}(A,A) & \ldots & P_{jk}(A,Y) \\ \relax
  \vdots & \ddots & \vdots \\ \relax
  P_{jk}(Y,A) & \ldots & P_{jk}(Y,Y)
 \end{pmatrix} \label{eq:eq5}
\end{equation}

\noindent where $j=1,2,\ldots,(n-1)$; $k=j+1,\ldots,n$.

We can also write,
\begin{equation}
 P_{jk}(a,b)=P_{jk}(a|b)p_k(b) \,, \label{eq:eq6}
\end{equation}
This equation can be also taken as another definition of joint
probability. $\!P\!_{jk}(\!a|b)$ is the Conditional probability of
occurrence of the amino acid \textbf{\emph{a}} in column
\textbf{\emph{j}} if the amino acid \textbf{\emph{b}} is already
found in column \textbf{\emph{k}}. We then have,
\begin{equation}
 \sum_a P_{jk}(a|b) = 1 \label{eq:eq7}
\end{equation}
From eqs.(\ref{eq:eq6}), (\ref{eq:eq7}), we have:
\begin{equation}
 \sum_a P_{jk}(a,b) = p_k(b) \label{eq:eq8}
\end{equation}
and from eq.(\ref{eq:eq8}),
\begin{equation}
 \sum_a\sum_b P_{jk}(a,b) = 1 \label{eq:eq9}
\end{equation}

\noindent which is an identity since $P_{jk}(a,b)$ is also a probability.

Eqs.(\ref{eq:eq8}) and (\ref{eq:eq9}) can be also derived from
\begin{equation}
 \sum_a n_{jk}(a,b) = n_k(b) \,;\quad \sum_a\sum_b n_{jk}(a,b)
 = m  \label{eq:eq10}
\end{equation}

\noindent and the definitions, eqs.(\ref{eq:eq1}), (\ref{eq:eq4}).

We now have from Bayes' law:
\begin{equation}
 P_{jk}(a|b)p_k(b) = P_{kj}(b|a)p_j(a) \label{eq:eq10a}
\end{equation}
and from eq.(\ref{eq:eq10a}), the property of symmetry,
\begin{equation}
 P_{jk}(a,b) = P_{kj}(b,a) \label{eq:eq10b}
\end{equation}

The matrices $P_{jk}$ can be organized in a triangular array:
\begin{equation}
 P =
 \begin{matrix}
  {} & P_{12} & P_{13} & P_{14} & \ldots & P_{1\,n-2} & P_{1\,n-1}
  & P_{1\,n} \\ \relax
  {} & {} & P_{23} & P_{24} & \ldots & P_{2\,n-2} & P_{2\,n-1}
  & P_{2\,n} \\ \relax
  {} & {} & {} & P_{34} & \ldots & P_{3\,n-2} & P_{3\,n-1}
  & P_{3\,n} \\ \relax
  {} & {} & {} & {} & \ddots & \vdots & \vdots & \vdots \\ \relax
  {} & {} & {} & {} & {} & P_{n-3\,n-2} & P_{n-3\,n-1}
  & P_{n-3\,n} \\ \relax
  {} & {} & {} & {} & {} & {} & P_{n-2\,n-1} & P_{n-2\,n}
  \\ \relax
  {} & {} & {} & {} & {} & {} & {} & P_{n-1\,n}
 \end{matrix} \label{eq:eq11}
\end{equation}
The number of matrices until the $P_{jk}$-th one is given by
\begin{equation}
 C_{jk} = j(n-1) - \frac{j(j-1)}{2} - (n-k) \label{eq:eq12}
\end{equation}

\noindent These numbers can be also arranged as a triangular array:
\begin{equation}
 \resizebox{.99\hsize}{!}{$C =
 \begin{matrix}
  1 & 2 & 3 & 4 & 5 & 6 & \ldots & (n-3) & (n-2) & (n-1) \\ \relax
  {} & {} & {} & {} & {} & {} & {} & {} & {} & {} \\ \relax
  {} & n & (n+1) & (n+2) & (n+3) & (n+4) & \ldots & (2n-5) & (2n-4) &
  (2n-3) \\ \relax
  {} & {} & {} & {} & {} & {} & {} & {} & {} & {} \\ \relax
  {} & {} & (2n-2) & (2n-1) & 2n & (2n+1) & \ldots & (3n-8) & (3n-7) &
  (3n-6) \\ \relax
  {} & {} & {} & {} & {} & {} & {} & {} & {} & {} \\ \relax
  {} & {} & {} & (3n-5) & (3n-4) & (3n-3) & \ldots & (4n-12) & (4n-11) &
  (4n-10) \\ \relax
  {} & {} & {} & {} & {} & {} & {} & {} & {} & {} \\ \relax
  {} & {} & {} & {} & (4n-9) & (4n-8) & \ldots & (5n-17) & (5n-16) & (5n-15)
  \\ \relax
  {} & {} & {} & {} & {} & {} & {} & {} & {} & {} \\ \relax
  {} & {} & {} & {} & {} & (5n-14) & \ldots & (6n-23) & (6n-22) & (6n-21)
  \\ \relax
  {} & {} & {} & {} & {} & {} & {} & {} & {} & {} \\ \relax
  {} & {} & {} & {} & {} & {} & \ddots & \vdots & \vdots & \vdots \\ \relax
  {} & {} & {} & {} & {} & {} & {} & {} & {} & {} \\ \relax
  {} & {} & {} & {} & {} & {} & {} & \frac{1}{2}(n^2-n-10) & \frac{1}{2}
  (n^2-n-8) & \frac{1}{2}(n+2)(n-3) \\ \relax
  {} & {} & {} & {} & {} & {} & {} & {} & {} & {} \\ \relax
  {} & {} & {} & {} & {} & {} & {} & {} & \frac{1}{2}(n^2-n-4) &
  \frac{1}{2}(n+1)(n-2) \\ \relax
  {} & {} & {} & {} & {} & {} & {} & {} & {} & {} \\ \relax
  {} & {} & {} & {} & {} & {} & {} & {} & {} & \frac{1}{2}\,n(n-1)
 \end{matrix}
 $} \label{eq:eq13}
\end{equation}

Eq.(\ref{eq:eq11}) should be used for the construction of a computational
code to perform all necessary calculations. We postpone to other publication
the presentation of some interesting results on the analysis of eq.(\ref{eq:eq13}).

The calculation of the matrix elements $P_{jk}(a,b)$ from a rectangular array
$m \times n$ of amino acids is done by the ``concatenation'' process
which is easily implemented on computational codes. We choose a pair of
columns $j=\bar{\jmath}$, $k=\bar{k}$ from the strings, $a=A$, $C$,
$\ldots$, $W$, $Y$, $b=A$, $C$, $\ldots$, $W$, $Y$ and we look for the
occurrence of the combinations $ab=AA$, $AC$, $\ldots$, $AW$, $AY$, $CA$,
$CC$, $\ldots$, $CW$, $CY$, $\ldots$, $WA$, $WC$, $\ldots$, $WW$, $WY$,
$\ldots$, $YA$, $YC$, $\ldots$, $YW$, $YY$. We then calculate their numbers
of occurrences $n_{\bar{\jmath}\bar{k}}(A,A)$,
$n_{\bar{\jmath}\bar{k}}(A,C)$, $\ldots$, $n_{\bar{\jmath}\bar{k}}(Y,W)$,
$n_{\bar{\jmath}\bar{k}}(Y,Y)$ and the corresponding probabilities
$P_{\bar{\jmath}\bar{k}}(A,A)$, $P_{\bar{\jmath}\bar{k}}(A,C)$, $\ldots$,\\
$P_{\bar{\jmath}\bar{k}}(Y,W)$, $P_{\bar{\jmath}\bar{k}}(Y,Y)$ from
eq.(\ref{eq:eq4}). We do the same for the other $\frac{n^2-n-2}{2}$ pairs of
columns.

As an example, let us suppose that we have the $3 \times 4$ array:
\begin{figure}[H]
 \centering
 \includegraphics[width=0.25\linewidth]{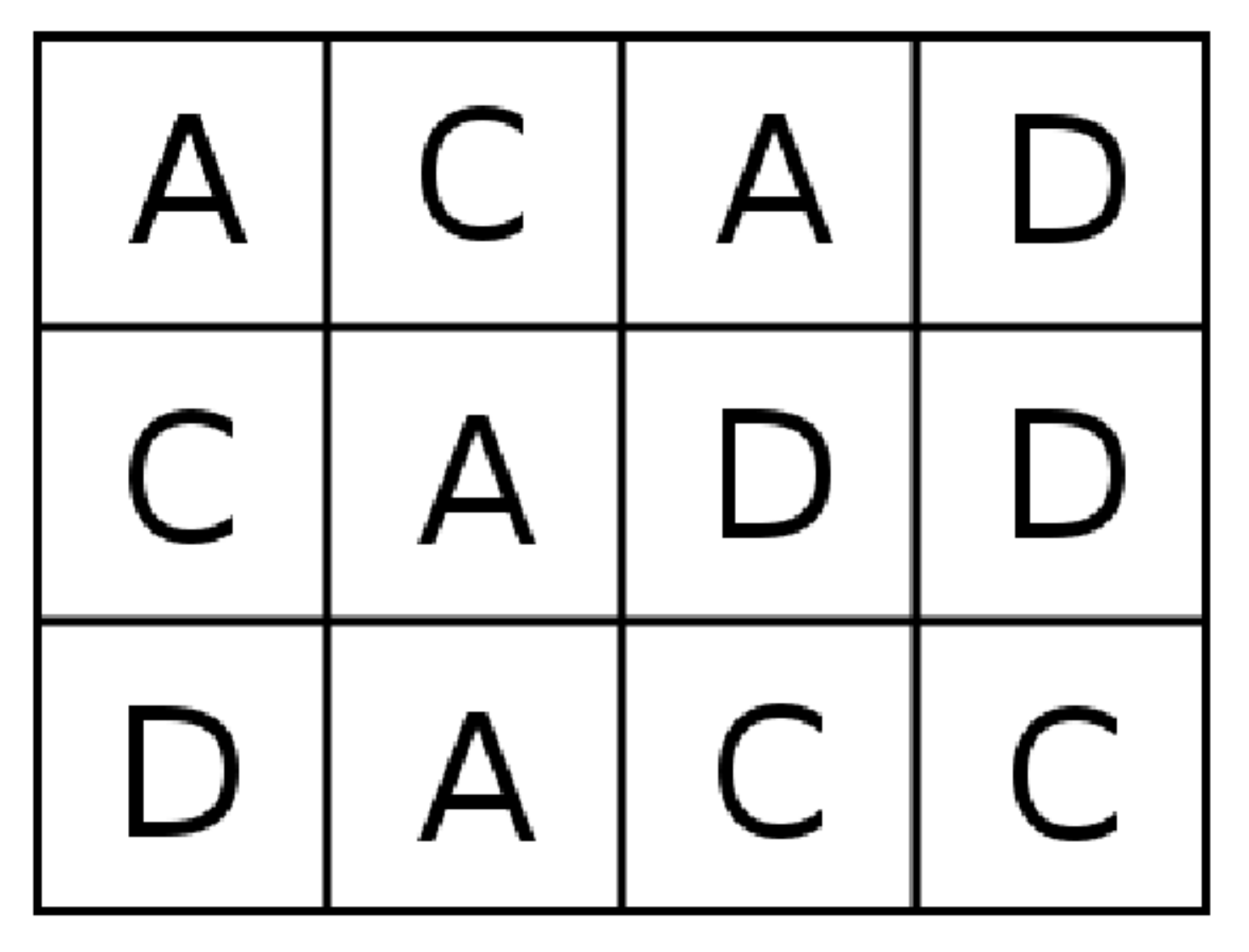}
 \caption{\small An example of a $3 \times 4$ array with amino acids
 A, C, D. \label{fig1}}
\end{figure}

Let us choose the pair of columns 1,2. We look for the occurrence of the
combinations $AA$, $AC$, $AD$, $CA$, $CC$, $CD$, $DA$, $DC$, $DD$ on the
pair of columns 1,2 of the array above and we found $n_{12}(A,C) = 1$,
$n_{12}(C,A) = 1$, $n_{12}(D,A) = 1$. The others $n_{12}(a,b) = 0$. From
eq.(\ref{eq:eq4}) we can write for the matrices $P_{jk}$ of eq.(\ref{eq:eq5}):
\begin{align}
 & P_{12} =
 \begin{pmatrix}
  0 & 1/3 & 0 \\ \relax
  1/3 & 0 & 0 \\ \relax
  1/3 & 0 & 0
 \end{pmatrix}
 \!;
 P_{13} =
 \begin{pmatrix}
  1/3 & 0 & 0 \\ \relax
  0 & 0 & 1/3 \\ \relax
  0 & 1/3 & 0
 \end{pmatrix}
 \!;
 P_{14} =
 \begin{pmatrix}
  0 & 0 & 1/3 \\ \relax
  0 & 0 & 1/3 \\ \relax
  0 & 1/3 & 0
 \end{pmatrix}
 \nonumber \\
 & P_{23} =
 \begin{pmatrix}
  0 & 1/3 & 1/3 \\ \relax
  1/3 & 0 & 0 \\ \relax
  0 & 0 & 0
 \end{pmatrix}
 \!;
 P_{24} =
 \begin{pmatrix}
  0 & 1/3 & 1/3 \\ \relax
  0 & 0 & 1/3 \\ \relax
  0 & 0 & 0
 \end{pmatrix}
 \!;
 P_{34} =
 \begin{pmatrix}
  0 & 0 & 1/3 \\ \relax
  0 & 1/3 & 0 \\ \relax
  0 & 0 & 1/3
 \end{pmatrix} \label{eq:eq14}
\end{align}
The Maple computing system ``recognizes'' the matricial structure through
its Linear Algebra package. The Perl computing system ``operates'' only
with ``strings''. The results above are easily obtained in Maple, but in
Perl we have to find alternative ways of calculating the joint probabilities.
The first method is to calculate the probabilities per row of the
$3 \times 4$ array. We have for the first row:
\begin{align}
 & \Pi_{12}^{(1)} =
 \begin{pmatrix}
  0 & 1/3 & 0 \\ \relax
  0 & 0 & 0 \\ \relax
  0 & 0 & 0
 \end{pmatrix}
 \!; \,\,
 \Pi_{13}^{(1)} =
 \begin{pmatrix}
  1/3 & 0 & 0 \\ \relax
  0 & 0 & 0 \\ \relax
  0 & 0 & 0
 \end{pmatrix}
 \!; \,\,
 \Pi_{14}^{(1)} =
 \begin{pmatrix}
  0 & 0 & 1/3 \\ \relax
  0 & 0 & 0 \\ \relax
  0 & 0 & 0
 \end{pmatrix}
 \nonumber \\
 & \Pi_{23}^{(1)} =
 \begin{pmatrix}
  0 & 0 & 0 \\ \relax
  1/3 & 0 & 0 \\ \relax
  0 & 0 & 0
 \end{pmatrix}
 \!; \,\,
 \Pi_{24}^{(1)} =
 \begin{pmatrix}
  0 & 0 & 0 \\ \relax
  0 & 0 & 1/3 \\ \relax
  0 & 0 & 0
 \end{pmatrix}
 \!; \,\,
 \Pi_{34}^{(1)} =
 \begin{pmatrix}
  0 & 0 & 1/3 \\ \relax
  0 & 0 & 0 \\ \relax
  0 & 0 & 0
 \end{pmatrix} \label{eq:eq15}
\end{align}
For the second row:
\begin{align}
 & \Pi_{12}^{(2)} =
 \begin{pmatrix}
  0 & 0 & 0 \\ \relax
  1/3 & 0 & 0 \\ \relax
  0 & 0 & 0
 \end{pmatrix}
 \!; \,\,
 \Pi_{13}^{(2)} =
 \begin{pmatrix}
  0 & 0 & 0 \\ \relax
  0 & 0 & 1/3 \\ \relax
  0 & 0 & 0
 \end{pmatrix}
 \!; \,\,
 \Pi_{14}^{(2)} =
 \begin{pmatrix}
  0 & 0 & 0 \\ \relax
  0 & 0 & 1/3 \\ \relax
  0 & 0 & 0
 \end{pmatrix}
 \nonumber \\
 & \Pi_{23}^{(2)} =
 \begin{pmatrix}
  0 & 0 & 1/3 \\ \relax
  0 & 0 & 0 \\ \relax
  0 & 0 & 0
 \end{pmatrix}
 \!; \,\,
 \Pi_{24}^{(2)} =
 \begin{pmatrix}
  0 & 0 & 1/3 \\ \relax
  0 & 0 & 0 \\ \relax
  0 & 0 & 0
 \end{pmatrix}
 \!; \,\,
 \Pi_{34}^{(2)} =
 \begin{pmatrix}
  0 & 0 & 0 \\ \relax
  0 & 0 & 0 \\ \relax
  0 & 0 & 1/3
 \end{pmatrix} \label{eq:eq16}
\end{align}
For the third row:
\begin{align}
 & \Pi_{12}^{(3)} =
 \begin{pmatrix}
  0 & 0 & 0 \\ \relax
  0 & 0 & 0 \\ \relax
  1/3 & 0 & 0
 \end{pmatrix}
 \!; \,\,
 \Pi_{13}^{(3)} =
 \begin{pmatrix}
  0 & 0 & 0 \\ \relax
  0 & 0 & 0 \\ \relax
  0 & 1/3 & 0
 \end{pmatrix}
 \!; \,\,
 \Pi_{14}^{(3)} =
 \begin{pmatrix}
  0 & 0 & 0 \\ \relax
  0 & 0 & 0 \\ \relax
  0 & 1/3 & 0
 \end{pmatrix}
 \nonumber \\
 & \Pi_{23}^{(3)} =
 \begin{pmatrix}
  0 & 1/3 & 0 \\ \relax
  0 & 0 & 0 \\ \relax
  0 & 0 & 0
 \end{pmatrix}
 \!; \,\,
 \Pi_{24}^{(3)} =
 \begin{pmatrix}
  0 & 1/3 & 0 \\ \relax
  0 & 0 & 0 \\ \relax
  0 & 0 & 0
 \end{pmatrix}
 \!; \,\,
 \Pi_{34}^{(3)} =
 \begin{pmatrix}
  0 & 0 & 0 \\ \relax
  0 & 1/3 & 0 \\ \relax
  0 & 0 & 0
 \end{pmatrix} \label{eq:eq17}
\end{align}

We stress that Perl does not recognize these matrix structures. This is just
our arrangement in order to make comparison with Maple calculations. However,
Perl ``knows'' how to sum the calculations done per rows to obtain:
\begin{equation}
 \Pi_{12}^{(1)}+\Pi_{12}^{(2)}+\Pi_{12}^{(3)} =
 \begin{pmatrix}
  0 & 1/3 & 0 \\ \relax
  1/3 & 0 & 0 \\ \relax
  1/3 & 0 & 0
 \end{pmatrix}
 \equiv P_{12} \label{eq:eq18}
\end{equation}
\begin{equation}
 \Pi_{13}^{(1)}+\Pi_{13}^{(2)}+\Pi_{13}^{(3)} =
 \begin{pmatrix}
  1/3 & 0 & 0 \\ \relax
  0 & 0 & 1/3 \\ \relax
  0 & 1/3 & 0
 \end{pmatrix}
 \equiv P_{13} \label{eq:eq19}
\end{equation}
\begin{equation}
 \Pi_{14}^{(1)}+\Pi_{14}^{(2)}+\Pi_{14}^{(3)} =
 \begin{pmatrix}
  0 & 0 & 1/3 \\ \relax
  0 & 0 & 1/3 \\ \relax
  0 & 1/3 & 0
 \end{pmatrix}
 \equiv P_{14} \label{eq:eq20}
\end{equation}
\begin{equation}
 \Pi_{23}^{(1)}+\Pi_{23}^{(2)}+\Pi_{23}^{(3)} =
 \begin{pmatrix}
  0 & 1/3 & 1/3 \\ \relax
  1/3 & 0 & 0 \\ \relax
  0 & 0 & 0
 \end{pmatrix}
 \equiv P_{23} \label{eq:eq21}
\end{equation}
\begin{equation}
 \Pi_{24}^{(1)}+\Pi_{24}^{(2)}+\Pi_{24}^{(3)} =
 \begin{pmatrix}
  0 & 1/3 & 1/3 \\ \relax
  0 & 0 & 1/3 \\ \relax
  0 & 0 & 0
 \end{pmatrix}
 \equiv P_{24} \label{eq:eq22}
\end{equation}
\begin{equation}
 \Pi_{34}^{(1)}+\Pi_{34}^{(2)}+\Pi_{34}^{(3)} =
 \begin{pmatrix}
  0 & 0 & 1/3 \\ \relax
  0 & 1/3 & 0 \\ \relax
  0 & 0 & 1/3
 \end{pmatrix}
 \equiv P_{34} \label{eq:eq23}
\end{equation}

We are then able to translate the Perl output in ``matrix language''.
However, this output does not come as an ordered set of joint
probabilities, as we have done by arranging the output in the form of the
matrices $\Pi_{jk}^{(l)}$, $j=1,2,3$, $k=2,3,4$, $l=1,2,3$. In order to
calculate functions of the probabilities as the entropy measures, it will
take too much time for the Perl computing system to collect the necessary
probability values. This is due to the ``Hash'' structure of the Perl as
compared to the usual ``array'' structure of the Maple. A new form of
arranging the strings to favour an a priori ordination will circumvent this
inconvenience of the ``hash'' structure. Let us then write the following
extended string associated to the $m \times n$ rectangular array:
\begin{equation*}
 \Big(\underbrace{(\overset{1}{A}\overset{2}{C}\overset{3}{D})}_{1}
 \underbrace{(\overset{1}{C}\overset{2}{A}\overset{3}{A})}_{2}
 \underbrace{(\overset{1}{A}\overset{2}{D}\overset{3}{C})}_{3}
 \underbrace{(\overset{1}{D}\overset{2}{D}\overset{3}{C})}_{4}\Big)
\end{equation*}
We then get
\begin{align}
 &P_{12}(A,C) = 1/3,\quad P_{12}(C,A) = 1/3,\quad P_{12}(D,A) = 1/3 \nonumber \\
 &P_{13}(A,A) = 1/3,\quad P_{13}(C,D) = 1/3,\quad P_{13}(D,C) = 1/3 \nonumber \\
 &P_{14}(A,D) = 1/3,\quad P_{14}(C,D) = 1/3,\quad P_{14}(D,C) = 1/3 \nonumber \\
 &P_{23}(C,A) = 1/3,\quad P_{23}(A,D) = 1/3,\quad P_{23}(A,C) = 1/3 \nonumber \\
 &P_{24}(C,D) = 1/3,\quad P_{24}(A,D) = 1/3,\quad P_{24}(A,C) = 1/3 \nonumber \\
 &P_{34}(A,D) = 1/3,\quad P_{34}(D,D) = 1/3,\quad P_{34}(C,C) = 1/3 \label{eq:eq24}
\end{align}
All the other joint probabilities $P_{jk}(a,b)$ are equal to zero.

This is a feasible treatment for the ``hash'' structure. In the example
solved above the probabilities will come already ordered in triads. This
will save time in the calculations with the Perl system.

It should be stressed that the Perl computing system does not recognize any
formal relations of Linear Algebra. However, it does quite well if these
relations are converted into products and sums. In order to give an example
of working with the Perl system, we take a calculation with the usual Shannon
Entropy measure. The calculation of the Entropy for the
columns $j$,$k$ is done by
\begin{equation}
 S_{jk} = -\sum_a\sum_b P_{jk}(a,b)\log P_{jk}(a,b) =
 -\mathrm{Tr}\Big(P_{jk}(\log P_{jk})^{\mathrm{T}}\Big) \label{eq:eq25}
\end{equation}

\noindent where $P_{jk}$ is the matrix given in eq.(\ref{eq:eq5}) and
$\mathrm{Tr,T}$ stands for the operations of taking the trace and transposing
a matrix, respectively. The matrix $(\log P_{jk})^{\mathrm{T}}$ is given by
\begin{equation*}
 (\log P_{jk})^{\mathrm{T}} =
 \begin{pmatrix}
  \log P_{jk}(A,A) & \ldots & \log P_{jk}(Y,A) \\ \relax
  \vdots & {} & \vdots \\ \relax
  \log P_{jk}(A,Y) & \ldots & \log P_{jk}(Y,Y)
 \end{pmatrix}
\end{equation*}

\noindent we also include for a useful reference, the matrix
\begin{equation*}
 p_j(p_k)^{\mathrm{T}} =
 \begin{pmatrix}
  p_j(A)p_k(A) & \ldots & p_j(A)p_k(Y) \\ \relax
  \vdots & {} & \vdots \\ \relax
  p_j(Y)p_k(A) & \ldots & p_j(Y)p_k(Y)
 \end{pmatrix}
\end{equation*}

Since from eqs.(\ref{eq:eq18})--(\ref{eq:eq23}), we have
\begin{equation}
 P_{jk} = \sum_{l=1}^m \Pi_{jk}^{(l)} \label{eq:eq26}
\end{equation}
we can write:
\begin{equation}
 S_{jk} = -\mathrm{Tr}\left(\left(\sum_{l=1}^m \Pi_{jk}^{(l)}\right)
 (\log P_{jk})^{\mathrm{T}}\right) = -\sum_{l=1}^m \mathrm{Tr}
 \Big(\Pi_{jk}^{(l)}(\log P_{jk})^{\mathrm{T}}\Big) \label{eq:eq27}
\end{equation}
There is no problem for calculating in Perl, if we prepare eq.(\ref{eq:eq26})
by expressing previously all the products and sums to be done. The real
problem with Perl calculations is the arrangement of the output of values
$P_{jk}(a,b)$, due to the ``hash'' structure as have been stressed above.

\section{Entropy Measures. The Sharma-Mittal set and the associated Jaccard
Entropy measure}
We start this section with the definition of the two-parameter Sharma-Mittal
entropies \cite{sharma,mondaini1,mondaini2}
\begin{align}
 (SM)_{jk}(r,s) &= -\frac{1}{1-r}\left(1-\left(\sum_a\sum_b\big(P_{jk}(a,b)
 \big)^s\right)^{\frac{1-r}{1-s}}\right) \label{eq:eq28} \\
 (SM)_{j}(r,s) &= -\frac{1}{1-r}\left(1-\left(\sum_a \big(p_{j}(a)
 \big)^s\right)^{\frac{1-r}{1-s}}\right) \label{eq:eq29}
\end{align}

\noindent where $p_j(a)$ and $P_{jk}(a,b)$ are the simple and joint probabilities
of occurrence of amino acids as defined on eqs.(\ref{eq:eq1}) and (\ref{eq:eq4}),
respectively. \textbf{\emph{r}}, \textbf{\emph{s}} are non-dimensional parameters.

We can associate to the entropy measures above their corresponding
one parameter forms to be given by the limits:
\begin{align}
 H_{jk}(s) &= \lim_{r \to s} (SM)_{jk}(r,s) = -\frac{1}{1-s}
 \left(1-\sum_a\sum_b\big(P_{jk}(a,b)\big)^s\right) \label{eq:eq30} \\
 H_{j}(s) &= \lim_{r \to s} (SM)_{j}(r,s) = -\frac{1}{1-s}
 \left(1-\sum_a\big(p_{j}(a)\big)^s\right) \label{eq:eq31}
\end{align}

\noindent These are the Havrda-Charvat Entropy Measures and they will be
specially emphasized in the present work. Other alternative proposals for
the single parameter entropies are given by

\noindent The Renyi's Entropy measures:
\begin{align}
 R_{jk}(s) &= \lim_{r \to 1} (SM)_{jk}(r,s) = \frac{1}{1-s}
 \log\left(\sum_a\sum_b\big(P_{jk}(a,b)\big)^s\right) \label{eq:eq32} \\
 R_{j}(s) &= \lim_{r \to 1} (SM)_{j}(r,s) = \frac{1}{1-s}
 \log\left(\sum_a\big(p_{j}(a)\big)^s\right) \label{eq:eq33}
\end{align}
The Landsberg-Vedral Entropy measures:
\begin{align}
 L_{jk}(s) =\! \lim_{\quad r \to 2-s} (SM)_{jk}(r,s) &= \frac{1}{1-s}
 \left(1-\Big(\sum_a\sum_b\big(P_{jk}(a,b)\big)^s\Big)^{-1}\right)
 \label{eq:eq34} \\
 &= \frac{H_{jk}(s)}{\sum\limits_a\sum\limits_b\big(P_{jk}(a,b)\big)^s}
 \nonumber \\
 L_{j}(s) =\! \lim_{\quad r \to 2-s} (SM)_{j}(r,s) &= \frac{1}{1-s}
 \left(1-\Big(\sum_a\big(p_{j}(a)\big)^s\Big)^{-1}\right) \label{eq:eq35} \\
 &= \frac{H_j(s)}{\sum\limits_a\big(p_j(a)\big)^s} \nonumber
\end{align}
All these Entropy measures have the free-parameter Shannon entropy
in the limit $s \to 1$.
\begin{align}
 \lim_{s \to 1} H_{jk}(s) &= \lim_{s \to 1} R_{jk}(s) = \lim_{s \to 1}
 L_{jk}(s) = S_{jk} \label{eq:eq36} \\
 \lim_{s \to 1} H_j(s) &= \lim_{s \to 1} R_j(s) = \lim_{s \to 1} L_j(s)
 = S_j \label{eq:eq37}
\end{align}
where
\begin{align}
 S_{jk} &= -\sum_a\sum_b P_{jk}(a,b)\log P_{jk}(a,b) \tag{\ref{eq:eq25}} \\
 S_j &= -\sum_a p_j(a)\log p_j(a) \label{eq:eq38}
\end{align}
are the Shannon entropy measures \cite{mondaini6}.

We now introduce a convenient version of a Mutual Information measure:
\begin{equation}
 M_{jk}(r,s) = \frac{1}{1-r}\left(1-\left(\frac{\sum\limits_a\sum\limits_b
 \big(P_{jk}(a,b)\big)^s}{\sum\limits_a\sum\limits_b\big(p_j(a)p_k(b)
 \big)^s}\right)^{\frac{1-r}{1-s}}\right) \label{eq:eq39}
\end{equation}

\noindent We can see that $M_{jk}(r,0) = 0$ and if $\exists\, \bar{\jmath}, \bar{k}$
such that $P_{\bar{\jmath}\bar{k}}(a,b) = p_{\bar{\jmath}}(a)p_{\bar{k}}(b)$
$\Rightarrow$ $M_{\bar{\jmath}\bar{k}}(r,s) = 0$. We also have,
\begin{equation}
 M_{jk}(1,s) = \lim_{r \to 1} M_{jk}(r,s) = -\frac{1}{1-s}\log\left(
 \frac{\sum\limits_a\sum\limits_b\big(P_{jk}(a,b)\big)^s}{\sum\limits_a
 \sum\limits_b\big(p_j(a)p_k(b)\big)^s}\right) \label{eq:eq40}
\end{equation}
and in the limit $s \to 1$
\begin{align}
 M_{jk} = \lim_{s \to 1} M_{jk}(1,s) =& \sum_a\sum_b P_{jk}(a,b)\log
 P_{jk}(a,b) \nonumber \\
 &- \sum_a\sum_b p_j(a)p_k(b)\log\big(p_j(a)p_k(b)\big) \label{eq:eq41}
\end{align}
and from the identities:
\begin{align*}
 &\sum_a p_j(a) = 1 \,,\, \forall j \,; \quad \sum_b p_k(b) = 1 \,,\,
 \forall k \\
 &\sum_a P_{jk}(a,b) = p_k(b) \,,\, \forall j \,; \quad \sum_b P_{jk}(a,b)
 = p_j(a) \,,\, \forall k
\end{align*}
obtained from eqs.(\ref{eq:eq3}), (\ref{eq:eq4}), (\ref{eq:eq6}), (\ref{eq:eq7}),
we can also write instead eq.(\ref{eq:eq41}):
\begin{equation}
 M_{jk} = \sum_a\sum_b P_{jk}(a,b)\log P_{jk}(a,b) - \sum_a\sum_b P_{jk}(a,b)
 \log\big(p_j(a)p_k(b)\big) \label{eq:eq42}
\end{equation}
It should be stressed that we are not assuming that $P_{jk}(a,b) \equiv
p_j(a)p_k(b)$ above. This equality is assumed to be valid only for
$j = \bar{\jmath}$, $k = \bar{k}$.

Eq.(\ref{eq:eq41}) or (\ref{eq:eq42}) can be also written as:
\begin{equation}
 M_{jk} = -S_{jk} + S_j + S_k \label{eq:eq43}
\end{equation}

\noindent where $S_{jk}$ and $S_j$, $S_k$ are the Shannon entropy measures for
joint and single probabilities, respectively, eqs.(\ref{eq:eq25}), (\ref{eq:eq38}).

As an additional topic, we emphasize that the Mutual Information measure can
be also derived from the Kullback-Leibler divergence \cite{mondaini6} which is
written as
\begin{equation}
 (KL)_{jk}(b) = \sum_a P_{jk}(a|b)\log\left(\frac{P_{jk}(a|b)}{p_j(a)}
 \right) \label{eq:eq44}
\end{equation}

\noindent where $P_{jk}(a|b)$ is the Conditional probability, eq.(\ref{eq:eq8}).
We then have,
\begin{equation}
 (KL)_{jk}(b) = \sum_a \frac{P_{jk}(a,b)}{p_k(b)}\log\left(\frac{P_{jk}
 (a,b)}{p_j(a)p_k(b)}\right) \label{eq:eq45}
\end{equation}

\noindent and the $M_{jk}$ mutual information measure will be given by
\begin{equation}
 M_{jk} = \sum_b p_k(b)(KL)_{jk}(b) = \sum_a\sum_b P_{jk}(a,b)\log\left(
 \frac{P_{jk}(a,b)}{p_j(a)p_k(b)}\right) \label{eq:eq46}
\end{equation}
which is the same as eq.(\ref{eq:eq42}), q.e.d.

As the last topic of this section, we now introduce the concept of
Information Distance and we then derive the Jaccard Entropy measure as
an obvious consequence. Let us write:
\begin{equation}
 d_{jk}(r,s) = H_{jk}(r,s) - M_{jk}(r,s) \label{eq:eq47}
\end{equation}
Since we are working with Entropy measures, we have to satisfy the
non-negativeness criteria:
\begin{equation}
 H_{jk}(r,s) \geq 0 \,;\quad M_{jk}(r,s) \geq 0 \,;\quad H_{jk}(r,s)
 -M_{jk}(r,s) \geq 0 \label{eq:eq48}
\end{equation}
This means that by satisfying the inequalities (\ref{eq:eq48}), restrictions
on the $r$, $s$ parameters should be discovered and considered for
the description of the protein databases by Entropy measures like $H_{jk}(r,s)$.

From inequalities (\ref{eq:eq48}), we can write,
\begin{equation}
 0 \leq d_{jk}(r,s) = H_{jk}(r,s) - M_{jk}(r,s) \leq H_{jk}(r,s)
 \label{eq:eq49}
\end{equation}
and
\begin{equation}
 0 \leq J_{jk}(r,s) \leq 1 \label{eq:eq50}
\end{equation}
where
\begin{equation}
 J_{jk}(r,s) = 1-\frac{M_{jk}(r,s)}{H_{jk}(r,s)} \label{eq:eq51}
\end{equation}

\noindent is the normalized Jaccard Entropy Measure as obtained from the
normalized Information Distance. We then give below the results of checking
the inequalities (\ref{eq:eq48}) for some families of the Pfam database. We
shall take the limit $r \to s$ and we work with the corresponding
one-parameter Entropy measures: $H_j(s)$, $H_{jk}(s)$, $M_{jk}(s)$,
$J_{jk}(s)$. We then have to check:
\begin{align*}
 H_{jk}(s) \geq 0 \,&,\quad M_{jk}(s) \geq 0 \,, \quad H_{jk}(s)-M_{jk}(s)
 \geq 0 \,, \\
 &0 \leq J_{jk}(s) = 1 - \frac{M_{jk}(s)}{H_{jk}(s)} \leq 1
\end{align*}
\begin{table}[H]
 \begin{center}
  \caption{\small Study of the non-negativeness of $H_{jk}(s)$, $M_{jk}(s)$
  and $d_{jk}(s)$ values for the protein family PF06850. \label{tab1}}
  \begin{tabular}{|c|c|c|c|}
   \hline
   $\mathbf{s}$ & $\mathbf{H_{jk}(s)}$ & $\mathbf{M_{jk}(s)}$ &
   $\mathbf{d_{jk}(s)}$ \\
   \hline
   $0.1$ & $0$ & $0$ & $0$ \\
   \hline
   $0.3$ & $0$ & $0$ & $0$ \\
   \hline
   $0.5$ & $0$ & $0$ & $0$ \\
   \hline
   $0.7$ & $0$ & $0$ & $0$ \\
   \hline
   $0.9$ & $0$ & $0$ & $0$ \\
   \hline
   $1.0$ & $0$ & $0$ & $0$ \\
   \hline
   $1.2$ & $0$ & $718$ & $16$ \\
   \hline
   $1.5$ & $0$ & $1708$ & $38$ \\
   \hline
   $1.7$ & $0$ & $2351$ & $61$ \\
   \hline
   $1.9$ & $0$ & $2898$ & $192$ \\
   \hline
   $2.0$ & $0$ & $3139$ & $309$ \\
   \hline
  \end{tabular}
 \end{center}
\end{table}

\noindent The $s$-values corresponding to negative $M_{jk}(s)$ values do not
lead to a useful characterization of the Jaccard Entropy measure according
to the inequality on eq.(\ref{eq:eq49}) which is violated in this case and
these $s$-values will not be taken into consideration. Other studies of the
Entropy values and specially those of the behaviour of the association of
entropies, will give additional restrictions on the feasible $s$-range. The
scope of the present work does not allow an intensive study of these
techniques of entropy association \cite{mondaini2} which will then appear on
a forthcoming contribution.

\begin{table}[!ht]
 \begin{center}
  \caption{\small Study of the non-negativeness of $H_{jk}(s)$, $M_{jk}(s)$
  and $d_{jk}(s)$ values for the protein family PF00135. \label{tab2}}
  \begin{tabular}{|c|c|c|c|}
   \hline
   $\mathbf{s}$ & $\mathbf{H_{jk}(s)}$ & $\mathbf{M_{jk}(s)}$ &
   $\mathbf{d_{jk}(s)}$ \\
   \hline
   $0.1$ & $0$ & $0$ & $0$ \\
   \hline
   $0.3$ & $0$ & $0$ & $0$ \\
   \hline
   $0.5$ & $0$ & $0$ & $0$ \\
   \hline
   $0.7$ & $0$ & $0$ & $0$ \\
   \hline
   $0.9$ & $0$ & $0$ & $0$ \\
   \hline
   $1.0$ & $0$ & $0$ & $0$ \\
   \hline
   $1.2$ & $0$ & $0$ & $0$ \\
   \hline
   $1.5$ & $0$ & $0$ & $467$ \\
   \hline
   $1.7$ & $0$ & $0$ & $14509$ \\
   \hline
   $1.9$ & $0$ & $0$ & $19026$ \\
   \hline
   $2.0$ & $0$ & $0$ & $19451$ \\
   \hline
  \end{tabular}
 \end{center}
\end{table}
\begin{table}[!hbt]
 \begin{center}
  \caption{\small Study of the non-negativeness of $H_{jk}(s)$, $M_{jk}(s)$
  and $d_{jk}(s)$ values for the protein family PF00005. \label{tab3}}
  \begin{tabular}{|c|c|c|c|}
   \hline
   $\mathbf{s}$ & $\mathbf{H_{jk}(s)}$ & $\mathbf{M_{jk}(s)}$ &
   $\mathbf{d_{jk}(s)}$ \\
   \hline
   $0.1$ & $0$ & $0$ & $0$ \\
   \hline
   $0.3$ & $0$ & $0$ & $0$ \\
   \hline
   $0.5$ & $0$ & $0$ & $0$ \\
   \hline
   $0.7$ & $0$ & $0$ & $0$ \\
   \hline
   $0.9$ & $0$ & $0$ & $0$ \\
   \hline
   $1.0$ & $0$ & $0$ & $0$ \\
   \hline
   $1.2$ & $0$ & $8$ & $5$ \\
   \hline
   $1.5$ & $0$ & $33$ & $4741$ \\
   \hline
   $1.7$ & $0$ & $55$ & $9679$ \\
   \hline
   $1.9$ & $0$ & $65$ & $12442$ \\
   \hline
   $2.0$ & $0$ & $69$ & $13203$ \\
   \hline
  \end{tabular}
 \end{center}
\end{table}

\begin{figure}[p]
 \centering
 \includegraphics[width=1\linewidth]{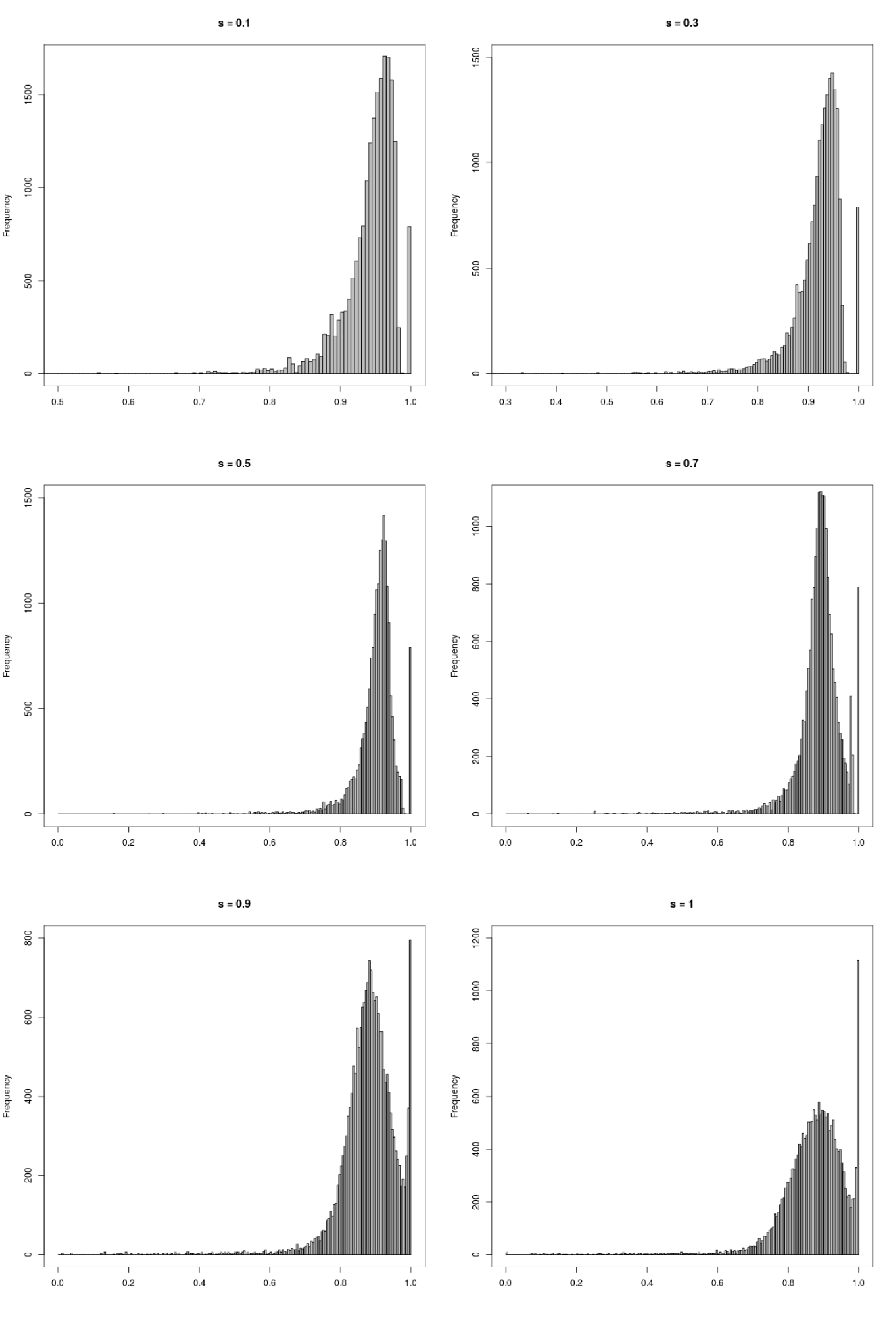}
 \caption{\small Histograms of Jaccard Entropy for family PF06850.
 \label{fig2}}
\end{figure}

\begin{figure}[p]
 \centering
 \includegraphics[width=1\linewidth]{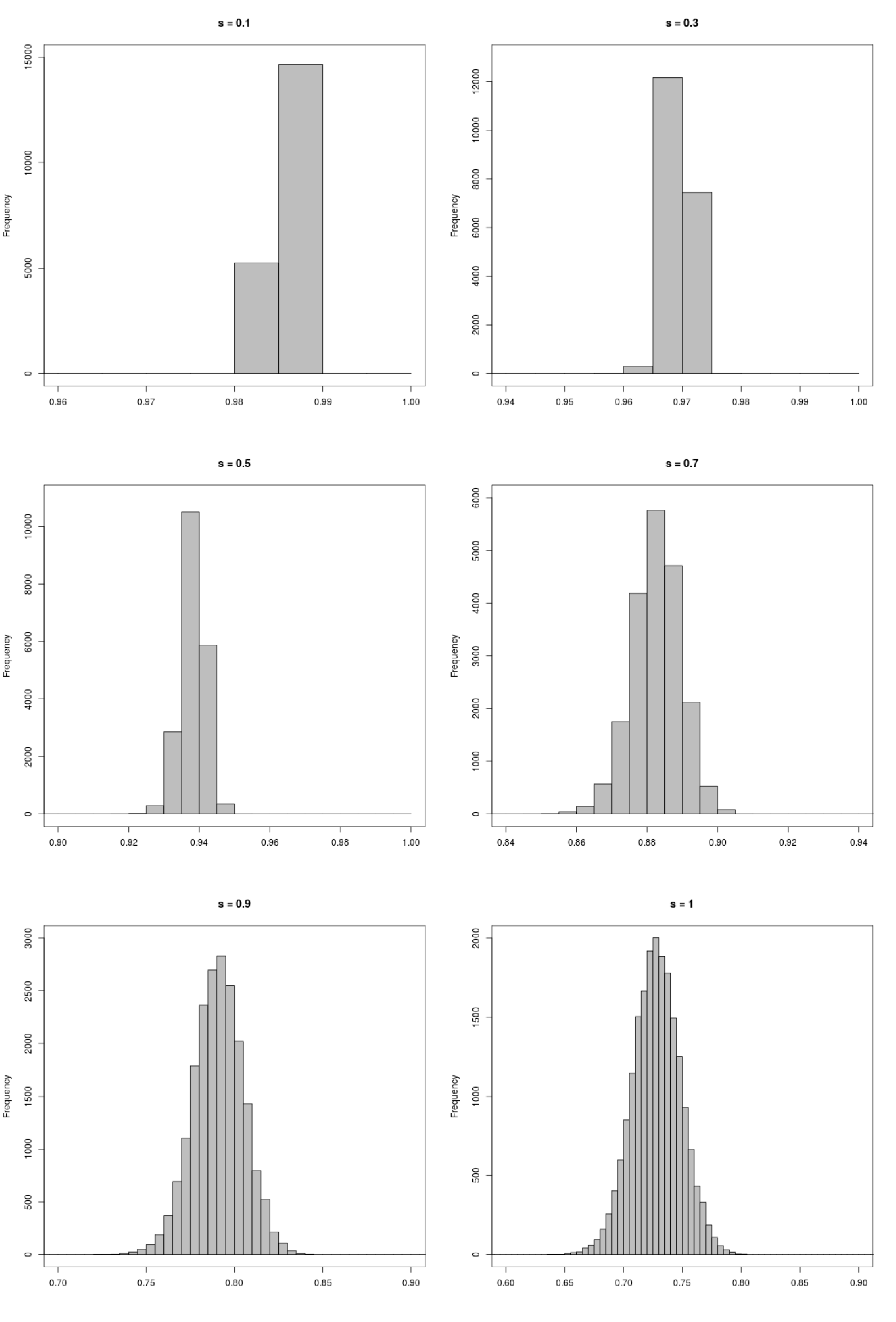}
 \caption{\small Histograms of Jaccard Entropy for family PF00135.
 \label{fig3}}
\end{figure}

\begin{figure}[p]
 \centering
 \includegraphics[width=1\linewidth]{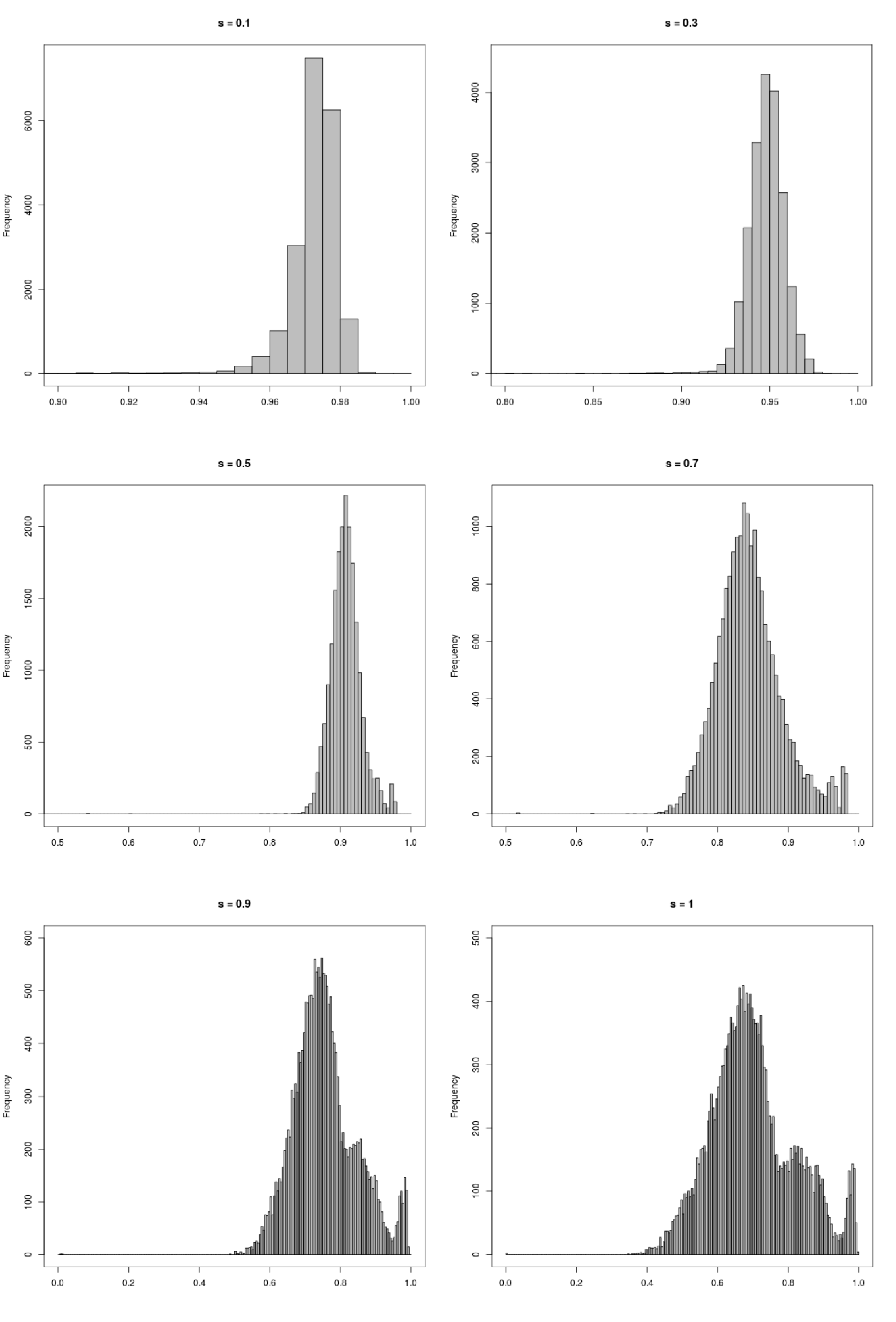}
 \caption{\small Histograms of Jaccard Entropy for family PF00005.
 \label{fig4}}
\end{figure}

The results on the previous three tables will clarify the idea of
restriction of the $s$-values of entropy measures for obtaining a
sound classification of families and clans on the Pfam database.
We now announce that the non-negativeness of the values of
$H_{jk}(s)$, $M_{jk}(s)$ and $d_{jk}(s)$ is actually guaranteed if
we restrict to $s \leq 1$ for all 1069 families which are classified
into 68 clans and already characterized at section 2. In figures
\ref{fig2}, \ref{fig3}, \ref{fig4}, we present the histograms of the
Jaccard Entropy measures for some $s \leq 1$ values of the
$s$-parameter.

\newpage

We also present the curves corresponding to the Average Jaccard Entropy
Measure (formula) for 09 families, a well-posed measure, with the
restriction $s \leq 1$, which is given by
\begin{equation*}
 J(s,f) = \frac{2}{n(n-1)}\sum_j\sum_k J_{jk}(s,f)
\end{equation*}

\begin{figure}[H]
 \centering
 \includegraphics[width=1\linewidth]{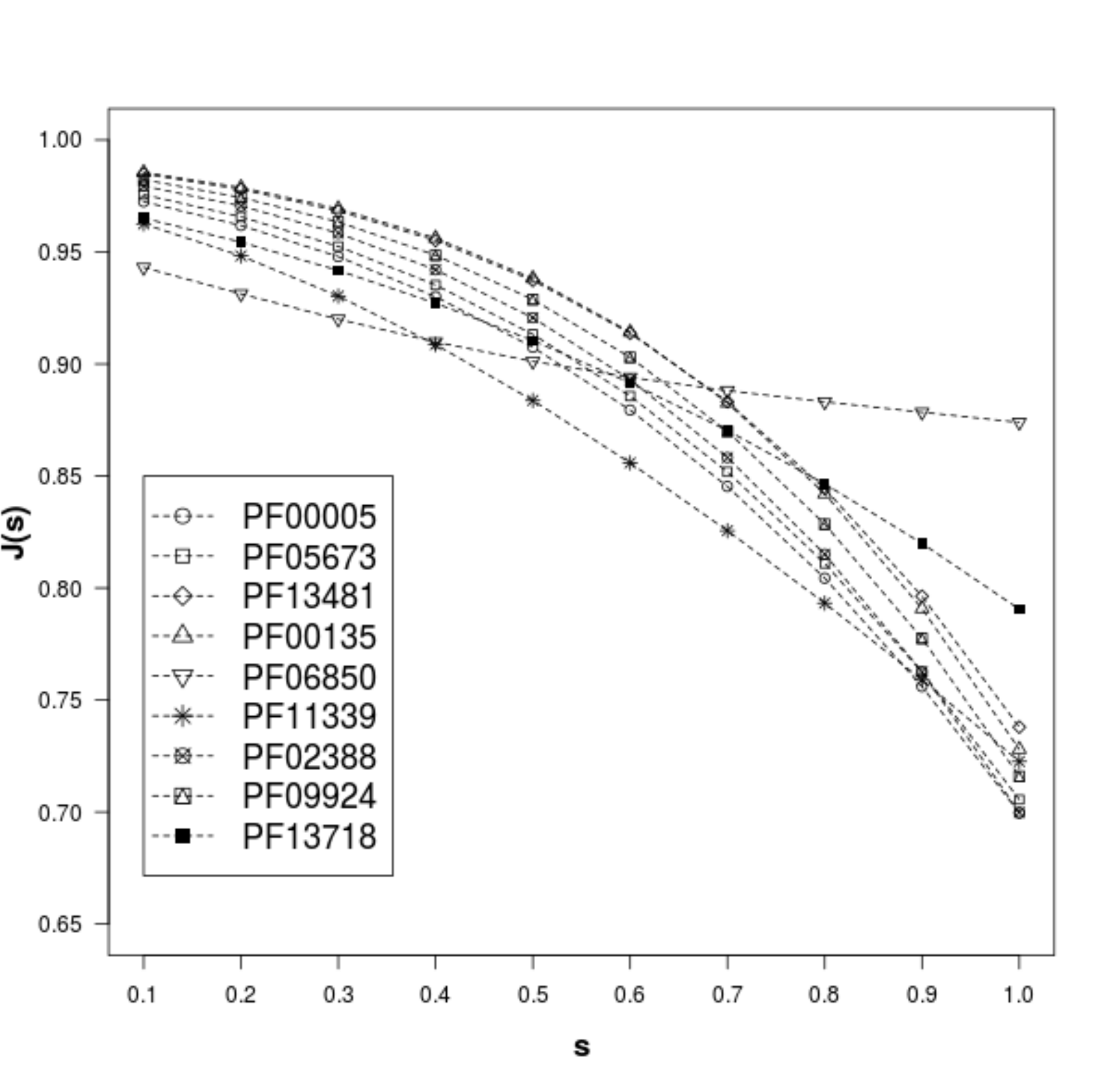}
 \caption{\small Curves of the Average Jaccard Entropy measures for families
 PF00005, PF05673, PF13481, PF00135, PF06850, PF11339, PF02388, PF09924,
 PF13718. \label{fig5}}
\end{figure}

\section{A First Assessment of Protein Databases with Entropy Measures}
As a motivation for future research to be developed in sections 6, 7, we
now introduce the first application of the formulae derived on the previous
sections in terms of a naive analysis of averages and standard deviations of
Entropy measure distributions. This will be also the first attempt at
classifying the distribution of amino acids in a generic protein database.
A robust approach to this research topic will be introduced and intensively
analyzed on sections 6, 7 with the introduction of ANOVA statistics and the
corresponding Hypothesis testing.

We then consider a Clan with \textbf{\emph{F}} families. The Havrda-Charvat
entropy measure associated to a pair of columns on the representative
$m \times n$ array of each family with a specified value of the
\textbf{\emph{s}} parameter is given by
\begin{equation}
 H_{jk}(s;f) = -\frac{1}{1-s} \left(1-\sum_a\sum_b\big(P_{jk}(a,b;f)
 \big)^s\right) \label{eq:eq52}
\end{equation}

\noindent We can then define an average of these entropy measures for
each family by
\begin{equation}
 \langle H(s;f) \rangle = \frac{2}{n(n-1)} \sum_j\sum_k H_{jk}(s;f)
 \label{eq:eq53}
\end{equation}

\noindent We also consider the average value of the averages over the
set of \emph{F} families:
\begin{equation}
 \langle H(s) \rangle_F = \frac{1}{F} \sum_{f=1}^F \langle H(s;f) \rangle
 \label{eq:eq54}
\end{equation}

\noindent The Standard deviation of the Entropy measures $H_{jk}(s;f)$
with relation to the given average in eq.(\ref{eq:eq53}) can be written
as:
\begin{equation}
 \sigma(s;f) = \left(\frac{1}{\frac{n(n-1)}{2}-1}\sum_j\sum_k\big(
 H_{jk}(s;f)-\langle H(s;f) \rangle\big)^2\right)^{1/2} \label{eq:eq55}
\end{equation}

\noindent and finally, the Standard deviation of the average
$\langle H(s;f)\rangle$ with respect to the average
$\langle H(s)\rangle_F$:
\begin{equation}
 \sigma_F(s) = \left(\frac{1}{F-1} \sum_{f=1}^F \big(\langle H(s;f)\rangle -
 \langle H(s)\rangle_F\big)^2\right)^{1/2} \label{eq:eq56}
\end{equation}

\begin{figure}[p]
 \centering
 \includegraphics[width=0.8\textwidth]{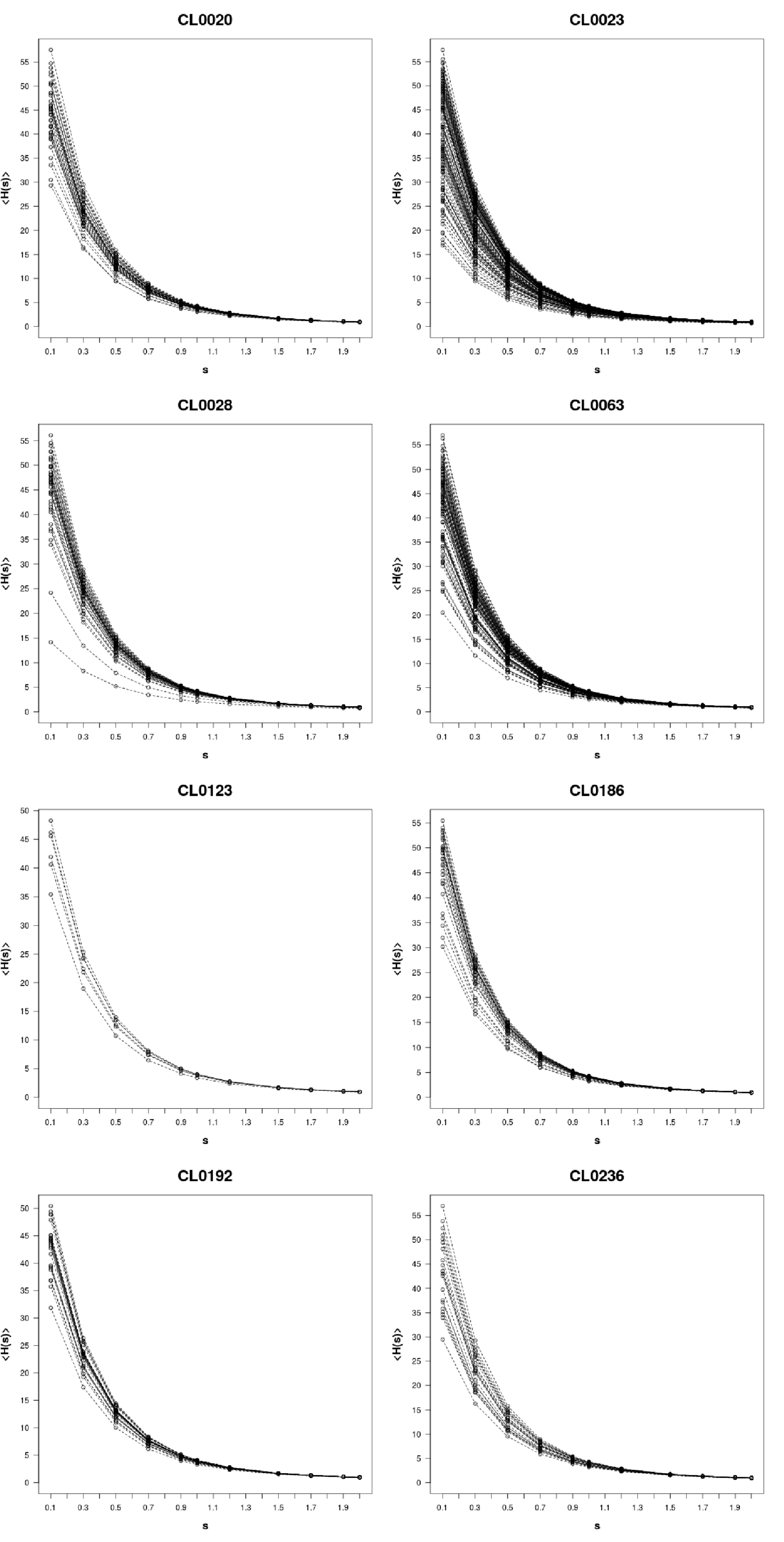}
 \caption{\small The Average values of the Havrda-Charvat Entropy measures
 for the families of a selected set of Clans and eleven values of the
 $s$-parameter, eq.(\ref{eq:eq53}). \label{fig6}}
\end{figure}

\begin{figure}[p]
 \centering
 \includegraphics[width=0.8\linewidth]{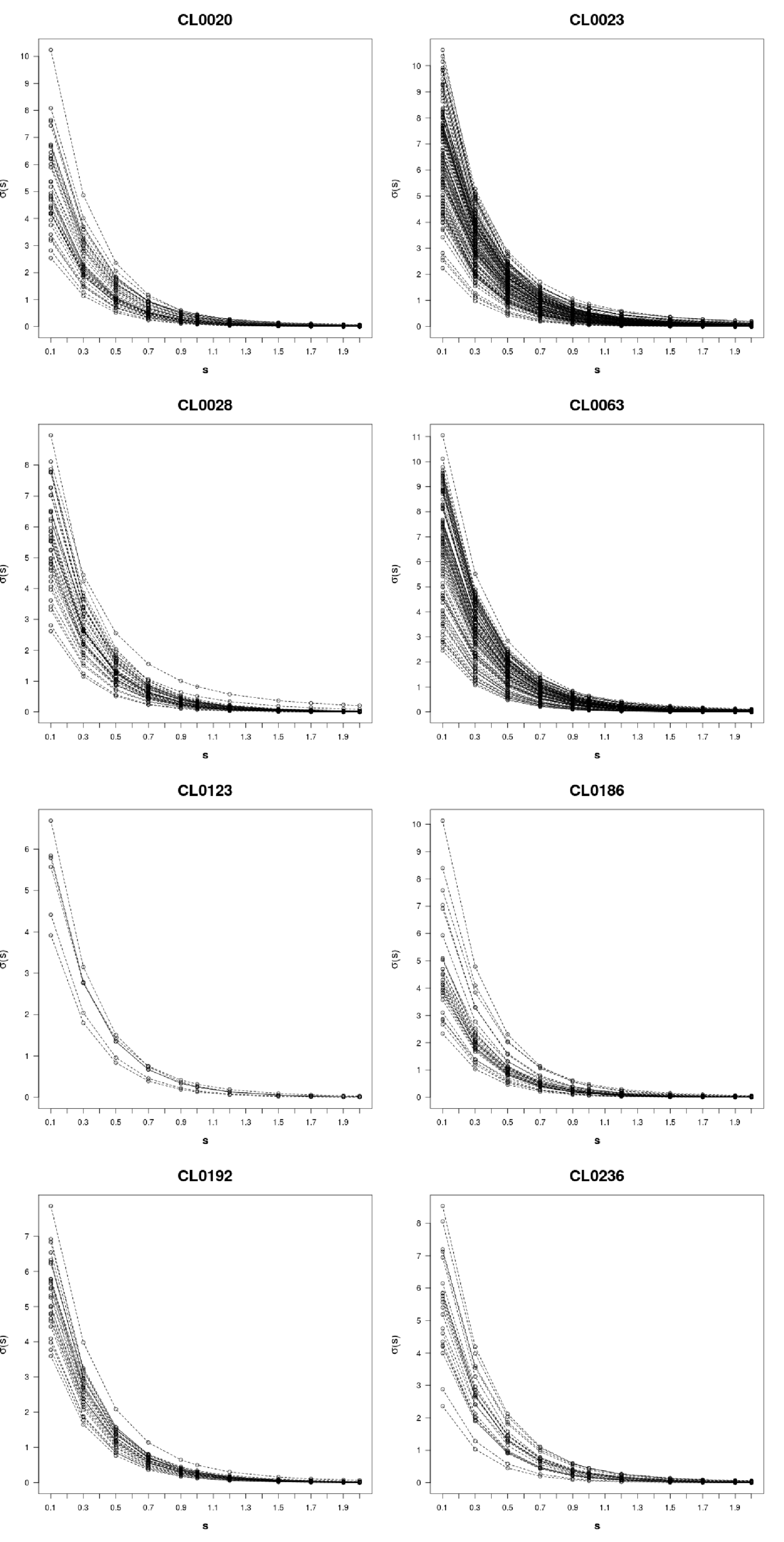}
 \caption{\small The standard deviation of the Havrda-Charvat Entropy
 measures with relation to the averages of these entropies for each family
 and eleven values of the  $s$-parameter, eq.(\ref{eq:eq55}). \label{fig7}}
\end{figure}

We present in figs.\ref{fig6}, \ref{fig7} below the diagrams corresponding to
formulae (\ref{eq:eq53}) and (\ref{eq:eq55}). We should stress that only
Clans with a minimum of five families are considered.
%
%
%
%

%

\newpage

We now present the values of $\langle H(s)\rangle_F$ and $\sigma_F(s)$ for
a selected number of Clans and eleven values of the $s$-parameter, according
to eqs.(\ref{eq:eq54}) and (\ref{eq:eq56}).
\begin{table}[H]
 \begin{center}
  \caption{\small The average values and the standard deviation of the Average
  Havrda-Charvat Entropy measures for eleven values of the $s$-parameter and
  a selected set of 8 Clans. \label{tab4}}
  \begingroup
  \everymath{\scriptstyle}
  \footnotesize
   \begin{tabular}{|c|c|c|c|c|c|c|c|c|}
    \hline
    \multicolumn{9}{|c|}{Clans from Pfam 27.0 --- Havrda-Charvat
    Entropies} \\
    \hline
    Clan number & $s$ & $\langle H(s)\rangle_F$ & $\sigma(s)_F$ & {} &
    Clan number & $s$ & $\langle H(s)\rangle_F$ & $\sigma(s)_F$ \\
    \cline{1-4} \cline{6-9}
    {} & $0.1$ & $44.212$ & $6.396$ & {} &
    {} & $0.1$ & $43.001$ & $4.667$ \\
    \cline{2-4} \cline{7-9}
    {} & $0.3$ & $23.375$ & $3.057$ & {} &
    {} & $0.3$ & $22.851$ & $2.295$ \\
    \cline{2-4} \cline{7-9}
    {} & $0.5$ & $12.992$ & $1.492$ & {} &
    {} & $0.5$ & $12.765$ & $1.161$ \\
    \cline{2-4} \cline{7-9}
    {} & $0.7$ & $7.645$ & $0.746$ & {} &
    {} & $0.7$ & $7.546$ & $0.605$ \\
    \cline{2-4} \cline{7-9}
    {} & $0.9$ & $4.784$ & $0.384$ & {} &
    {} & $0.9$ & $4.741$ & $0.326$ \\
    \cline{2-4} \cline{7-9}
    CL0020 & $1.0$ & $3.875$ & $0.278$ & {} &
    CL0123 & $1.0$ & $3.846$ & $0.242$ \\
    \cline{2-4} \cline{7-9}
    (38 families) & $1.2$ & $2.661$ & $0.150$ & {} &
    (06 families) & $1.2$ & $2.648$ & $0.137$ \\
    \cline{2-4} \cline{7-9}
    {} & $1.5$ & $1.680$ & $0.063$ & {} &
    {} & $1.5$ & $1.676$ & $0.062$ \\
    \cline{2-4} \cline{7-9}
    {} & $1.7$ & $1.311$ & $0.038$ & {} &
    {} & $1.7$ & $1.309$ & $0.038$ \\
    \cline{2-4} \cline{7-9}
    {} & $1.9$ & $1.062$ & $0.023$ & {} &
    {} & $1.9$ & $1.061$ & $0.024$ \\
    \cline{2-4} \cline{7-9}
    {} & $2.0$ & $0.967$ & $0.018$ & {} &
    {} & $2.0$ & $0.966$ & $0.019$ \\
    \cline{1-4} \cline{6-9}
    {} & $0.1$ & $39.235$ & $10.175$ & {} &
    {} & $0.1$ & $46.084$ & $6.790$ \\
    \cline{2-4} \cline{7-9}
    {} & $0.3$ & $20.908$ & $4.984$ & {} &
    {} & $0.3$ & $24.260$ & $3.224$ \\
    \cline{2-4} \cline{7-9}
    {} & $0.5$ & $11.733$ & $2.510$ & {} &
    {} & $0.5$ & $13.417$ & $1.561$ \\
    \cline{2-4} \cline{7-9}
    {} & $0.7$ & $6.982$ & $1.305$ & {} &
    {} & $0.7$ & $7.853$ & $0.772$ \\
    \cline{2-4} \cline{7-9}
    {} & $0.9$ & $4.422$ & $0.704$ & {} &
    {} & $0.9$ & $4.886$ & $0.392$ \\
    \cline{2-4} \cline{7-9}
    CL0023 & $1.0$ & $3.604$ & $0.525$ & {} &
    CL0186 & $1.0$ & $3.949$ & $0.282$ \\
    \cline{2-4} \cline{7-9}
    (119 families) & $1.2$ & $2.504$ & $0.302$ & {} &
    (29 families) & $1.2$ & $2.701$ & $0.149$ \\
    \cline{2-4} \cline{7-9}
    {} & $1.5$ & $1.606$ & $0.144$ & {} &
    {} & $1.5$ & $1.696$ & $0.060$ \\
    \cline{2-4} \cline{7-9}
    {} & $1.7$ & $1.263$ & $0.093$ & {} &
    {} & $1.7$ & $1.320$ & $0.035$ \\
    \cline{2-4} \cline{7-9}
    {} & $1.9$ & $1.030$ & $0.063$ & {} &
    {} & $1.9$ & $1.067$ & $0.020$ \\
    \cline{2-4} \cline{7-9}
    {} & $2.0$ & $0.940$ & $0.053$ & {} &
    {} & $2.0$ & $0.970$ & $0.016$ \\
    \cline{1-4} \cline{6-9}
    {} & $0.1$ & $44.906$ & $7.996$ & {} &
    {} & $0.1$ & $42.862$ & $4.566$ \\
    \cline{2-4} \cline{7-9}
    {} & $0.3$ & $23.671$ & $3.906$ & {} &
    {} & $0.3$ & $22.791$ & $2.190$ \\
    \cline{2-4} \cline{7-9}
    {} & $0.5$ & $13.114$ & $1.961$ & {} &
    {} & $0.5$ & $12.741$ & $1.072$ \\
    \cline{2-4} \cline{7-9}
    {} & $0.7$ & $7.692$ & $1.015$ & {} &
    {} & $0.7$ & $7.538$ & $0.537$ \\
    \cline{2-4} \cline{7-9}
    {} & $0.9$ & $4.799$ & $0.545$ & {} &
    {} & $0.9$ & $4.739$ & $0.275$ \\
    \cline{2-4} \cline{7-9}
    CL0028 & $1.0$ & $3.882$ & $0.406$ & {} &
    CL0192 & $1.0$ & $3.847$ & $0.199$ \\
    \cline{2-4} \cline{7-9}
    (41 families) & $1.2$ & $2.660$ & $0.232$ & {} &
    (26 families) & $1.2$ & $2.650$ & $0.107$ \\
    \cline{2-4} \cline{7-9}
    {} & $1.5$ & $1.676$ & $0.109$ & {} &
    {} & $1.5$ & $1.678$ & $0.044$ \\
    \cline{2-4} \cline{7-9}
    {} & $1.7$ & $1.307$ & $0.070$ & {} &
    {} & $1.7$ & $1.311$ & $0.026$ \\
    \cline{2-4} \cline{7-9}
    {} & $1.9$ & $1.058$ & $0.047$ & {} &
    {} & $1.9$ & $1.062$ & $0.015$ \\
    \cline{2-4} \cline{7-9}
    {} & $2.0$ & $0.963$ & $0.039$ & {} &
    {} & $2.0$ & $0.967$ & $0.012$ \\
    \cline{1-4} \cline{6-9}
    {} & $0.1$ & $42.312$ & $8.023$ & {} &
    {} & $0.1$ & $43.251$ & $7.469$ \\
    \cline{2-4} \cline{7-9}
    {} & $0.3$ & $22.454$ & $3.857$ & {} &
    {} & $0.3$ & $22.905$ & $3.564$ \\
    \cline{2-4} \cline{7-9}
    {} & $0.5$ & $12.534$ & $1.896$ & {} &
    {} & $0.5$ & $12.757$ & $1.734$ \\
    \cline{2-4} \cline{7-9}
    {} & $0.7$ & $7.411$ & $0.956$ & {} &
    {} & $0.7$ & $7.524$ & $0.863$ \\
    \cline{2-4} \cline{7-9}
    {} & $0.9$ & $4.660$ & $0.496$ & {} &
    {} & $0.9$ & $4.719$ & $0.440$ \\
    \cline{2-4} \cline{7-9}
    CL0063 & $1.0$ & $3.784$ & $0.362$ & {} &
    CL0236 & $1.0$ & $3.828$ & $0.317$ \\
    \cline{2-4} \cline{7-9}
    (92 families) & $1.2$ & $2.611$ & $0.198$ & {} &
    (21 families) & $1.2$ & $2.636$ & $0.169$ \\
    \cline{2-4} \cline{7-9}
    {} & $1.5$ & $1.658$ & $0.086$ & {} &
    {} & $1.5$ & $1.669$ & $0.069$ \\
    \cline{2-4} \cline{7-9}
    {} & $1.7$ & $1.297$ & $0.051$ & {} &
    {} & $1.7$ & $1.304$ & $0.040$ \\
    \cline{2-4} \cline{7-9}
    {} & $1.9$ & $1.053$ & $0.032$ & {} &
    {} & $1.9$ & $1.058$ & $0.024$ \\
    \cline{2-4} \cline{7-9}
    {} & $2.0$ & $0.960$ & $0.026$ & {} &
    {} & $2.0$ & $0.964$ & $0.019$ \\
    \cline{1-4} \cline{6-9}
   \end{tabular}
   \endgroup
  \end{center}
\end{table}

\begin{figure}[!hbt]
 \centering
 \includegraphics[width=1\linewidth]{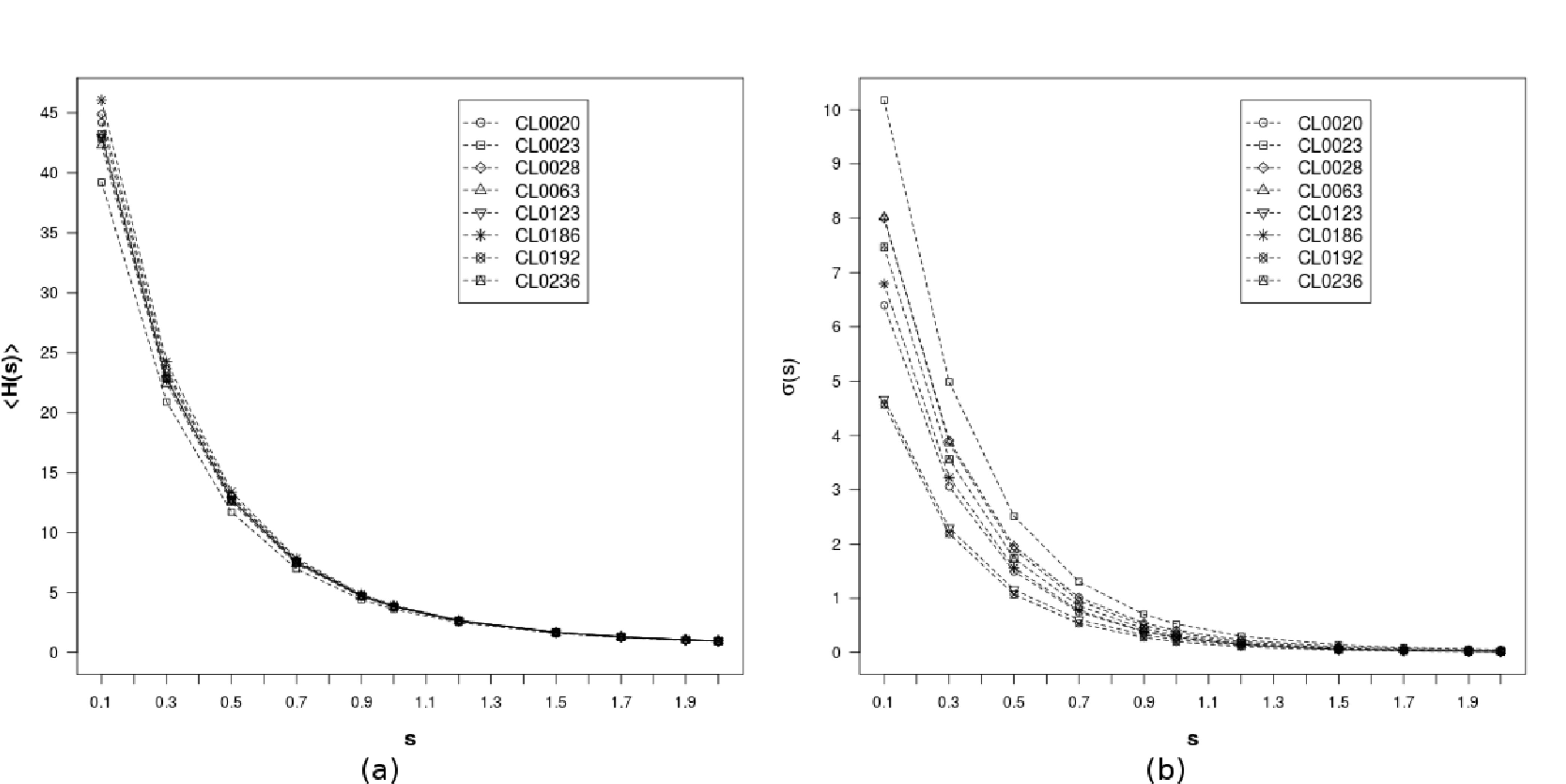}
 \caption{\small (a) The average values of Havrda-Charvat Entropies for a
 set of 08 Clans. (b) The Standard Deviation of the averages for
 Havrda-Charvat Entropies for a set of 08 Clans.\label{fig8}}
\end{figure}

In figs.\ref{fig8}a and \ref{fig8}b below, we present the graphs
corresponding to Table \ref{tab4}. These results just point out a more
elaborate formulation of the problem.

We now proceed to analyze a proposal (naive) for testing the robustness of
the Clan concept. We will check if Pseudo-Clans which have the same number
of families (a minimum of 05 families) of the corresponding Clans will have
essentially different values of $\langle H(s)\rangle_F$ and $\sigma_F(s)$.
The families to be associated with a Pseudo-Clan are obtained by sorting on
the set of 1069 families and by withdrawal of the families already sorted.
In Table \ref{tab5} below we present the values $\langle H(s)\rangle_F$ and
$\sigma_F(s)$ for the Pseudo-Clans obtained by the procedure described above.

The Figures 9a, 9b, do correspond to the comparison of data of Table
\ref{tab4} (Clans) with those of Table \ref{tab5} (Pseudo-Clans). Clans are
in red, Pseudo-Clans in blue.
\begin{figure}[H]
 \centering
 \includegraphics[width=1\linewidth]{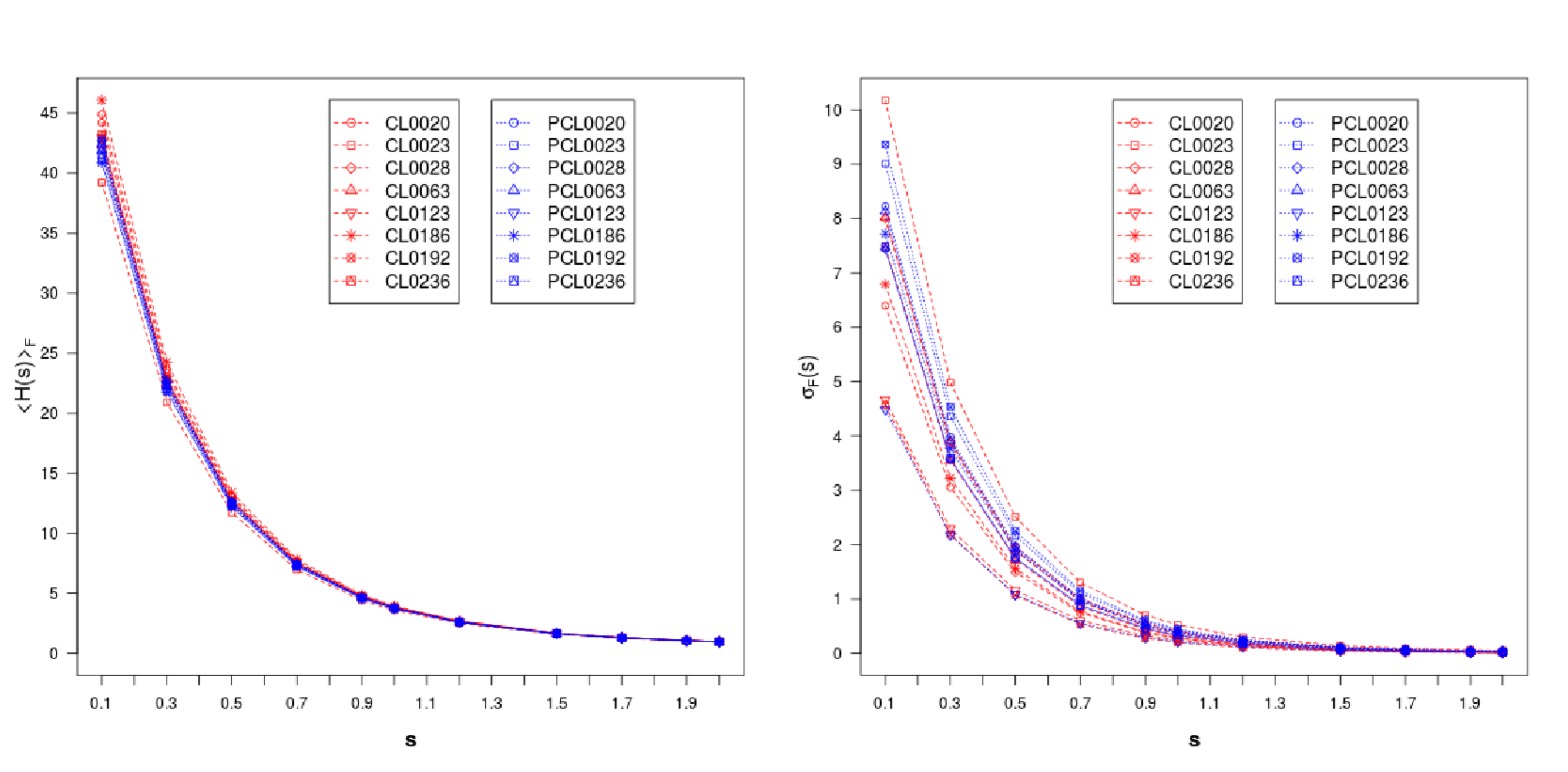}
 \caption{\small (a) The comparison of Clans and Pseudo-Clans average values.
 (b) The comparison of Clans and Pseudo-Clans standard deviation values.
 \label{fig9}}
\end{figure}
\begin{table}[!ht]
 \begin{center}
  \caption{\small The average values and the standard deviation of the
  Average Havrda-Charvat Entropy measures for eleven values of the
  $s$-parameter and a selected set of 8 Pseudo-Clans. \label{tab5}}
  \begingroup
  \everymath{\scriptstyle}
  \footnotesize
   \begin{tabular}{|c|c|c|c|c|c|c|c|c|}
    \hline
    \multicolumn{9}{|c|}{Pseudo-Clans / Pfam 27.0 --- Havrda-Charvat
    Entropies} \\
    \hline
    Pseudo-Clan & \multicolumn{1}{c|}{\multirow{2}{*}{$s$}} &
    \multicolumn{1}{c|}{\multirow{2}{*}{$\langle H(s)\rangle_F$}} &
    \multicolumn{1}{c|}{\multirow{2}{*}{$\sigma(s)_F$}} &
    \multicolumn{1}{c|}{} & Pseudo-Clan &
    \multicolumn{1}{c|}{\multirow{2}{*}{$s$}} &
    \multicolumn{1}{c|}{\multirow{2}{*}{$\langle H(s)\rangle_F$}} &
    \multicolumn{1}{c|}{\multirow{2}{*}{$\sigma(s)_F$}} \\
    number & \multicolumn{1}{c|}{} & \multicolumn{1}{c|}{} &
    \multicolumn{1}{c|}{} & \multicolumn{1}{c|}{} & number &
    \multicolumn{1}{c|}{} & \multicolumn{1}{c|}{} &
    \multicolumn{1}{c|}{} \\
    \cline{1-4} \cline{6-9}
    {} & $0.1$ & $42.381$ & $8.224$ & {} &
    {} & $0.1$ & $42.042$ & $4.484$ \\
    \cline{2-4} \cline{7-9}
    {} & $0.3$ & $22.480$ & $3.979$ & {} &
    {} & $0.3$ & $22.359$ & $2.175$ \\
    \cline{2-4} \cline{7-9}
    {} & $0.5$ & $12.542$ & $1.972$ & {} &
    {} & $0.5$ & $12.508$ & $1.082$ \\
    \cline{2-4} \cline{7-9}
    {} & $0.7$ & $7.410$ & $1.005$ & {} &
    {} & $0.7$ & $7.409$ & $0.555$ \\
    \cline{2-4} \cline{7-9}
    {} & $0.9$ & $4.657$ & $0.528$ & {} &
    {} & $0.9$ & $4.667$ & $0.293$ \\
    \cline{2-4} \cline{7-9}
    PCL0020 & $1.0$ & $3.781$ & $0.389$ & {} &
    PCL0123 & $1.0$ & $3.791$ & $0.216$ \\
    \cline{2-4} \cline{7-9}
    (38 families) & $1.2$ & $2.607$ & $0.216$ & {} &
    (06 families) & $1.2$ & $2.618$ & $0.120$ \\
    \cline{2-4} \cline{7-9}
    {} & $1.5$ & $1.655$ & $0.096$ & {} &
    {} & $1.5$ & $1.663$ & $0.053$ \\
    \cline{2-4} \cline{7-9}
    {} & $1.7$ & $1.295$ & $0.059$ & {} &
    {} & $1.7$ & $1.301$ & $0.032$ \\
    \cline{2-4} \cline{7-9}
    {} & $1.9$ & $1.051$ & $0.037$ & {} &
    {} & $1.9$ & $1.056$ & $0.020$ \\
    \cline{2-4} \cline{7-9}
    {} & $2.0$ & $0.958$ & $0.030$ & {} &
    {} & $2.0$ & $0.962$ & $0.016$ \\
    \cline{1-4} \cline{6-9}
    {} & $0.1$ & $41.023$ & $9.007$ & {} &
    {} & $0.1$ & $40.888$ & $7.719$ \\
    \cline{2-4} \cline{7-9}
    {} & $0.3$ & $21.825$ & $4.360$ & {} &
    {} & $0.3$ & $21.790$ & $3.768$ \\
    \cline{2-4} \cline{7-9}
    {} & $0.5$ & $12.219$ & $2.163$ & {} &
    {} & $0.5$ & $12.220$ & $1.891$ \\
    \cline{2-4} \cline{7-9}
    {} & $0.7$ & $7.247$ & $1.104$ & {} &
    {} & $0.7$ & $7.258$ & $0.981$ \\
    \cline{2-4} \cline{7-9}
    {} & $0.9$ & $4.573$ & $0.582$ & {} &
    {} & $0.9$ & $4.585$ & $0.529$ \\
    \cline{2-4} \cline{7-9}
    PCL0023 & $1.0$ & $3.714$ & $0.427$ & {} &
    PCL0186 & $1.0$ & $3.730$ & $0.394$ \\
    \cline{2-4} \cline{7-9}
    (119 families) & $1.2$ & $2.573$ & $0.240$ & {} &
    (29 families) & $1.2$ & $2.582$ & $0.227$ \\
    \cline{2-4} \cline{7-9}
    {} & $1.5$ & $1.640$ & $0.109$ & {} &
    {} & $1.5$ & $1.646$ & $0.109$ \\
    \cline{2-4} \cline{7-9}
    {} & $1.7$ & $1.286$ & $0.068$ & {} &
    {} & $1.7$ & $1.290$ & $0.071$ \\
    \cline{2-4} \cline{7-9}
    {} & $1.9$ & $1.046$ & $0.044$ & {} &
    {} & $1.9$ & $1.048$ & $0.049$ \\
    \cline{2-4} \cline{7-9}
    {} & $2.0$ & $0.954$ & $0.037$ & {} &
    {} & $2.0$ & $0.956$ & $0.041$ \\
    \cline{1-4} \cline{6-9}
    {} & $0.1$ & $42.716$ & $7.435$ & {} &
    {} & $0.1$ & $42.756$ & $9.361$ \\
    \cline{2-4} \cline{7-9}
    {} & $0.3$ & $23.658$ & $3.573$ & {} &
    {} & $0.3$ & $22.666$ & $4.535$ \\
    \cline{2-4} \cline{7-9}
    {} & $0.5$ & $12.641$ & $1.756$ & {} &
    {} & $0.5$ & $12.635$ & $2.253$ \\
    \cline{2-4} \cline{7-9}
    {} & $0.7$ & $7.468$ & $0.886$ & {} &
    {} & $0.7$ & $7.458$ & $1.151$ \\
    \cline{2-4} \cline{7-9}
    {} & $0.9$ & $4.692$ & $0.460$ & {} &
    {} & $0.9$ & $4.681$ & $0.608$ \\
    \cline{2-4} \cline{7-9}
    PCL0028 & $1.0$ & $3.808$ & $0.335$ & {} &
    PCL0192 & $1.0$ & $3.797$ & $0.448$ \\
    \cline{2-4} \cline{7-9}
    (41 families) & $1.2$ & $2.625$ & $0.183$ & {} &
    (26 families) & $1.2$ & $2.615$ & $0.250$ \\
    \cline{2-4} \cline{7-9}
    {} & $1.5$ & $1.664$ & $0.079$ & {} &
    {} & $1.5$ & $1.657$ & $0.113$ \\
    \cline{2-4} \cline{7-9}
    {} & $1.7$ & $1.302$ & $0.048$ & {} &
    {} & $1.7$ & $1.296$ & $0.070$ \\
    \cline{2-4} \cline{7-9}
    {} & $1.9$ & $1.056$ & $0.030$ & {} &
    {} & $1.9$ & $1.051$ & $0.046$ \\
    \cline{2-4} \cline{7-9}
    {} & $2.0$ & $0.962$ & $0.024$ & {} &
    {} & $2.0$ & $0.958$ & $0.037$ \\
    \cline{1-4} \cline{6-9}
    {} & $0.1$ & $41.870$ & $8.134$ & {} &
    {} & $0.1$ & $41.551$ & $7.476$ \\
    \cline{2-4} \cline{7-9}
    {} & $0.3$ & $22.252$ & $3.914$ & {} &
    {} & $0.3$ & $22.090$ & $3.591$ \\
    \cline{2-4} \cline{7-9}
    {} & $0.5$ & $12.440$ & $1.929$ & {} &
    {} & $0.5$ & $12.357$ & $1.763$ \\
    \cline{2-4} \cline{7-9}
    {} & $0.7$ & $7.366$ & $0.977$ & {} &
    {} & $0.7$ & $7.322$ & $0.887$ \\
    \cline{2-4} \cline{7-9}
    {} & $0.9$ & $4.638$ & $0.510$ & {} &
    {} & $0.9$ & $4.614$ & $0.459$ \\
    \cline{2-4} \cline{7-9}
    PCL0063 & $1.0$ & $3.771$ & $0.375$ & {} &
    PCL0236 & $1.0$ & $3.752$ & $0.334$ \\
    \cline{2-4} \cline{7-9}
    (92 families) & $1.2$ & $2.603$ & $0.207$ & {} &
    (21 families) & $1.2$ & $2.595$ & $0.181$ \\
    \cline{2-4} \cline{7-9}
    {} & $1.5$ & $1.654$ & $0.092$ & {} &
    {} & $1.5$ & $1.652$ & $0.077$ \\
    \cline{2-4} \cline{7-9}
    {} & $1.7$ & $1.295$ & $0.057$ & {} &
    {} & $1.7$ & $1.294$ & $0.046$ \\
    \cline{2-4} \cline{7-9}
    {} & $1.9$ & $1.052$ & $0.036$ & {} &
    {} & $1.9$ & $1.051$ & $0.028$ \\
    \cline{2-4} \cline{7-9}
    {} & $2.0$ & $0.959$ & $0.030$ & {} &
    {} & $2.0$ & $0.959$ & $0.022$ \\
    \cline{1-4} \cline{6-9}
   \end{tabular}
   \endgroup
  \end{center}
\end{table}

From the figures and tables above we can see that the region $s \leq 1$ leads
to a better characterization of the Entropy measures distributions on protein
databases.

For completeness, we list some useful formulae obtained from
eqs.(\ref{eq:eq53}), (\ref{eq:eq54}), (\ref{eq:eq56}) which help to predict
the profile of the curves above:
\begin{equation}
 \langle H(s;f) \rangle = -\frac{1}{1-s} \left(1-\frac{2}{n(n-1)}
 \sum_{a,b,j,k} e^{-s \lvert \log P_{jk}(a,b;f)\rvert}\right) \label{eq:eq57}
\end{equation}
\begin{equation}
 \langle H(s) \rangle_F = -\frac{1}{1-s} \left(1-\frac{2}{Fn(n-1)}
 \sum_{f,a,b,j,k} e^{-s \lvert \log P_{jk}(a,b;f)\rvert} \right)
 \label{eq:eq58}
\end{equation}
\begin{equation}
 \sigma_F(s) = \frac{2(F-1)^{1/2}}{Fn(n-1)}\left(\sum_{f=1}^F\left(\sum_{a,b,
 j,k} e^{-s \lvert \log P_{jk}(a,b;f)\rvert}\right)^2\right)^{1/2}
 \label{eq:eq59}
\end{equation}

\section{The treatment of data with the Maple Computing system and its
inadequacy for calculating Joint probabilities of occurrence. Alternative
systems.}
In this section, we specifically study the performance of the Maple system and
an example of alternative computing system, the Perl system, for calculating
the simple and joint probabilities of occurrences of amino acids. We also use
these two systems for calculating 19 $s$-power values of these probabilities.
We now select the family PF06850 in order to get an idea of the CPU and real
times which are necessary for calculating the probabilities and their powers
for the set of 1069 families. We start the calculation by adopting the Maple
system version 18. There are some comments to be made on the construction of
a computational code for calculating joint probabilities. This will be done
in detail at the end of CPU and real times for the calculation of the simple
and joint probabilities by using the developed code. The table below will
repeat the times for calculating $200 \times 20 = 4 \times 10^3$ and $200
\times \frac{200-1}{2} \times 20 \times 20 = 7.96 \times 10^6$ of simple and
joint probability values, respectively, for the PF06850 Pfam family.

\begin{table}[!ht]
 \begin{center}
  \caption{\small CPU time and real times for the calculation of the simple and
  joint probabilities of occurrence associated with the protein family
  PF06850. \label{tab6}}
  \begin{tabular}{|c|c|c|}
   \hline
   Maple System, version 18 & $t_{CPU}$ (sec) & $t_R$ (sec) \\
   \hline
   Simple probabilities & $0.527$ & $0.530$ \\
   \hline
   Joint probabilities & $5073.049$ & $4650.697$ \\
   \hline
  \end{tabular}
 \end{center}
\end{table}

After calculating all values of the probabilities $p_j(a)$ and $P_{jk}(a,b)$,
we can proceed to evaluate the powers $\big(p_j(a)\big)^s$ and $\big(P_jk(a,
b)\big)^s$ for 19 $s$-values. Our aim will be to use these values for
calculating the Entropy Measures according to eqs.(\ref{eq:eq30}),
(\ref{eq:eq31}). It should be noticed that the values of $p_j(a)$ and
$P_{jk}(a,b)$, have to be calculated only once by using a specific
computational code already referred on this work. Nevertheless, the use of
the code for calculating the joint probabilities associated to 1069 protein
families is the hardest of all calculations to be undertaken and it takes
too much time. These probabilities once calculated should be grouped in sets
of 400 values each corresponding to a pair of columns $j$, $k$ among the
$\frac{n(n-1)}{2}$ feasible ones and the calculating of
entropy value associated to this pair of columns $j$, $k$.
Given a $s$-value and after calculating the entropy of this
first pair $H_{1\,2}$ as a function of the 400 variables
$\big(P_{jk}(a,b)\big)^s$, $j \neq 1$, $k \neq 2$ and he/she will proceeds to
calculate again all values of $\frac{n(n-1)}{2} \times (20)^2$ in order
to extract another value of joint probability for calculating the
corresponding entropy value. This seems to be associated to
the unknowing of the concepts of a function of several variables,
unfortunately. After circumventing these mistakes coming from a bad
educational formation, we succeed at keeping all calculated values of the
probabilities and we then proceed to the calculation of the powers $\big(
p_j(a)\big)^s$, $\big(P_{jk}(a,b)\big)^s$ of these values and the
corresponding entropy measures. In tables \ref{tab7}, \ref{tab8}, \ref{tab9},
\ref{tab10} below, we report all these calculations for 19 values of the
$s$-parameter.

\begin{table}[!ht]
 \begin{center}
  \caption{\small CPU and real times for the calculation of 19 $s$-powers
  of simple probabilities of occurrence associated with the protein family
  PF06850. \label{tab7}}
  \begin{tabular}{|c|c|c|}
   \hline
   \multicolumn{3}{|c|}{Maple System, version 18,} \\
   \multicolumn{3}{|c|}{$s$-powers of probability $\big(p_j(a)\big)^s$}\\
   \hline
   $s$ & $t_{CPU}$ (sec) & $t_R$ (sec) \\
   \hline
   $0.1$ & $0.263$ & $0.358$ \\
   \hline
   $0.2$ & $0.137$ & $0.145$ \\
   \hline
   $0.3$ & $0.268$ & $0.277$ \\
   \hline
   $0.4$ & $0.139$ & $0.153$ \\
   \hline
   $0.5$ & $0.240$ & $0.219$ \\
   \hline
   $0.6$ & $0.144$ & $0.157$ \\
   \hline
   $0.7$ & $0.276$ & $0.254$ \\
   \hline
   $0.8$ & $0.144$ & $0.157$ \\
   \hline
   $0.9$ & $0.264$ & $0.235$ \\
   \hline
   $1.0$ & $0.088$/$0.151$ & $0.095$/$0.307$ \\
   \hline
   $2.0$ & $0.153$ & $0.095$ \\
   \hline
   $3.0$ & $0.128$ & $0.131$ \\
   \hline
   $4.0$ & $0.148$ & $0.141$ \\
   \hline
   $5.0$ & $0.096$ & $0.144$ \\
   \hline
   $6.0$ & $0.148$ & $0.167$ \\
   \hline
   $7.0$ & $0.148$ & $0.155$ \\
   \hline
   $8.0$ & $0.181$ & $0.094$ \\
   \hline
   $9.0$ & $0.104$ & $0.092$ \\
   \hline
   $10.0$ & $0.104$ & $0.100$ \\
   \hline
   Total & $3.173$ & $3.164$ \\
   \hline
  \end{tabular}
 \end{center}
\end{table}

\begin{table}[!ht]
 \begin{center}
  \caption{\small CPU and real times for the calculation of 19 $s$-powers
  of joint probabilities of occurrence associated with the protein family
  PF06850. \label{tab8}}
  \begin{tabular}{|c|c|c|}
   \hline
   \multicolumn{3}{|c|}{Maple System, version 18,} \\
   \multicolumn{3}{|c|}{$s$-powers of probability $\big(P_{jk}(a,b)\big)^s$}
   \\
   \hline
   $s$ & $t_{CPU}$ (sec) & $t_R$ (sec) \\
   \hline
   $0.1$ & $390.432$ & $206.646$ \\
   \hline
   $0.2$ & $382.887$ & $202.282$ \\
   \hline
   $0.3$ & $401.269$ & $210.791$ \\
   \hline
   $0.4$ & $416.168$ & $216.993$ \\
   \hline
   $0.5$ & $427.572$ & $221.541$ \\
   \hline
   $0.6$ & $430.604$ & $223.227$ \\
   \hline
   $0.7$ & $421.904$ & $218.484$ \\
   \hline
   $0.8$ & $434.888$ & $224.267$ \\
   \hline
   $0.9$ & $431.948$ & $223.023$ \\
   \hline
   $1.0$ & $442.933$/$482.612$ & $224.731$/$259.301$ \\
   \hline
   $2.0$ & $176.212$ & $147.455$ \\
   \hline
   $3.0$ & $234.100$ & $174.853$ \\
   \hline
   $4.0$ & $289.184$ & $181.552$ \\
   \hline
   $5.0$ & $327.740$ & $178.117$ \\
   \hline
   $6.0$ & $334.800$ & $194.691$ \\
   \hline
   $7.0$ & $349.064$ & $195.258$ \\
   \hline
   $8.0$ & $361.304$ & $195.437$ \\
   \hline
   $9.0$ & $386.217$ & $197.150$ \\
   \hline
   $10.0$ & $397.276$ & $197.868$ \\
   \hline
   Total & $7036.502$ & $3834.366$ \\
   \hline
  \end{tabular}
 \end{center}
\end{table}

The last row in tables \ref{tab7}, \ref{tab8}, includes the times necessary
for calculating the probabilities of table \ref{tab6}.

We are then able to proceed to the calculation of the corresponding
Havrda-Charvat entropy measures, $H_j(s)$, $H_{jk}(s)$: The results for 19
$s$-values are given in tables \ref{tab9}, \ref{tab10} below.

\begin{table}[!ht]
 \begin{center}
  \caption{\small CPU and real times for the calculation of the Entropy measures
  $H_j(s)$ for the protein family PF06850. \label{tab9}}
  \begin{tabular}{|c|c|c|}
   \hline
   \multicolumn{3}{|c|}{Maple System, version 18,} \\
   \multicolumn{3}{|c|}{Entropy Measures $H_j(s)$}\\
   \hline
   $s$ & $t_{CPU}$ (sec) & $t_R$ (sec) \\
   \hline
   $0.1$ & $0.148$ & $0.261$ \\
   \hline
   $0.2$ & $0.084$ & $0.153$ \\
   \hline
   $0.3$ & $0.120$ & $0.189$ \\
   \hline
   $0.4$ & $0.124$ & $0.198$ \\
   \hline
   $0.5$ & $0.160$ & $0.299$ \\
   \hline
   $0.6$ & $0.092$ & $0.139$ \\
   \hline
   $0.7$ & $0.159$ & $0.199$ \\
   \hline
   $0.8$ & $0.137$ & $0.175$ \\
   \hline
   $0.9$ & $0.120$ & $0.166$ \\
   \hline
   $1.0$ & $0.192$/$0.175$ & $0.339$/$0.159$ \\
   \hline
   $2.0$ & $0.144$ & $0.099$ \\
   \hline
   $3.0$ & $0.147$ & $0.105$ \\
   \hline
   $4.0$ & $0.084$ & $0.101$ \\
   \hline
   $5.0$ & $0.136$ & $0.070$ \\
   \hline
   $6.0$ & $0.096$ & $0.119$ \\
   \hline
   $7.0$ & $0.115$ & $0.078$ \\
   \hline
   $8.0$ & $0.120$ & $0.109$ \\
   \hline
   $9.0$ & $0.133$ & $0.080$ \\
   \hline
   $10.0$ & $0.132$ & $0.133$ \\
   \hline
   Total & $2.443$ & $3.012$ \\
   \hline
  \end{tabular}
 \end{center}
\end{table}

\begin{table}[!hb]
 \begin{center}
  \caption{\small CPU and real times for the calculation of the Entropy measures
  $H_{jk}(s)$ for the protein family PF06850. \label{tab10}}
  \begin{tabular}{|c|c|c|}
   \hline
   \multicolumn{3}{|c|}{Maple System, version 18,} \\
   \multicolumn{3}{|c|}{Entropy Measures $H_{jk}(s)$}\\
   \hline
   $s$ & $t_{CPU}$ (sec) & $t_R$ (sec) \\
   \hline
   $0.1$ & $156.332$ & $133.242$ \\
   \hline
   $0.2$ & $160.797$ & $136.706$ \\
   \hline
   $0.3$ & $169.024$ & $140.960$ \\
   \hline
   $0.4$ & $176.824$ & $147.853$ \\
   \hline
   $0.5$ & $184.120$ & $150.163$ \\
   \hline
   $0.6$ & $190.304$ & $154.058$ \\
   \hline
   $0.7$ & $196.633$ & $157.750$ \\
   \hline
   $0.8$ & $205.940$ & $164.101$ \\
   \hline
   $0.9$ & $215.559$ & $169.549$ \\
   \hline
   $1.0$ & $253.648$/$124.501$ & $204.634$/$148.993$ \\
   \hline
   $2.0$ & $141.148$ & $184.030$ \\
   \hline
   $3.0$ & $158.536$ & $167.173$ \\
   \hline
   $4.0$ & $173.136$ & $181.282$ \\
   \hline
   $5.0$ & $197.680$ & $238.723$ \\
   \hline
   $6.0$ & $215.000$ & $111.476$ \\
   \hline
   $7.0$ & $145.257$ & $115.221$ \\
   \hline
   $8.0$ & $156.848$ & $122.957$ \\
   \hline
   $9.0$ & $157.300$ & $126.233$ \\
   \hline
   $10.0$ & $166.399$ & $135.080$ \\
   \hline
   Total & $3420.485$ & $2941.791$ \\
   \hline
  \end{tabular}
 \end{center}
\end{table}

The total time for calculating all the Havrda-Charvat Entropy measure
content of probabilities of occurrence of amino acids on a
specific family is given in table \ref{tab11} below.

\begin{table}[!hb]
 \begin{center}
  \caption{\small Total CPU and real times for calculating the Entropy measure
  content of a family PF06850 and approximations for Grand
  Total of all sample space. \label{tab11}}
  \begingroup
  \everymath{\scriptstyle}
  \small
  \begin{tabular}{|c|c|c|}
   \hline
   Maple System, & Entropy Measures &
   Entropy Measures \\
   version 18 & $H\!_j(s)$ --- 19 $s$-values & $H\!_{jk}(s)$ ---
   19 $s$-values \\
   \hline
   Total CPU time & $0.527+3.173+2.443$
   & $5,073.049+7,036.502+$ \\
   (family PF06850) & $=6.143$ sec & $3,420.485=15,530.036$ sec \\
   \hline
   Grand Total CPU time & $6,566.867$ sec
   & $16,601,608.484$ sec \\
   (1069 families) & $=1.824$ hs & $=192.148$ days \\
   \hline
   Total Real time & $0.530+3.164+3.012$
   & $4,650.697+3,834.366+$ \\
   (family PF06850) & $=6.706$ sec & $2,941.791=11,426.854$ sec \\
   \hline
   Grand Total Real time & $7,168.714$ sec
   & $12,215,306.926$ sec \\
   (1069 families) & $=1.991$ hs & $=141.381$ days \\
   \hline
  \end{tabular}
  \endgroup
 \end{center}
\end{table}

From inspections of table \ref{tab11}, we realize that the Total CPU
and real times for calculating the Havrda-Charvat entropies $H_j(s)$, of
simple probabilities of occurrence $p_j(a)$ are obtained by summing up the
total time results from tables \ref{tab6}, \ref{tab7} and \ref{tab9}. For the
Havrda-Charvat entropies $H_{jk}(s)$, we have to sum up the total times at table
\ref{tab6}, \ref{tab8} and \ref{tab10}. We take for granted that the times for
calculating the Entropy Measure content of each family will not differ too much
and the results 3rd and 5th rows of table \ref{tab11} are obtained by multiplying
by 1069 --- the number of families in the sample space.

The results of table \ref{tab11} suggest the inadequacy of the Maple computing
system for analyzing the Entropy measure content of an example of protein database.
We have restricted ourselves to operate with usual operating systems, Linux
or OSX, on laptops. We have also worked with the alternative Perl computing
system. The Maple computing system has an ``array'' structure which is very
effective for doing calculations which require a knowledge of mathematical
methods. On the contrary, the alternative Perl computing system has a
``hash'' structure as was emphasized in the 2nd section and it operates very
well elementary operations with very large numbers. It is essential the
comparison of a senior erudite which is largely conversant with a large amount
of mathematical methods versus a ``genius'' brought to fame by media,
who is able only to multiply in a very fast way, numbers of many digits.

In order to specify the probabilities of computational configurations with
the usual desktops and laptops, we list below some of them which have been
used in the present work. The computing systems were the Maple ($M$) and Perl
($P$), the operating systems, the Linux ($L$) and Mac OSX ($O$) and the
structures: The Array I ($A_I$), Array II ($A_{I\!I}$) and Hash ($H$) (page 8,
section 2). The available computational configurations to undertake the task
of assessment of protein databases with Entropy measures could be listed as:
\begin{enumerate}
 \item $MLA_I$ --- Maple, Linux, Array I
 \item $POH$ --- Perl, OSX, Hash
 \item $PLA_{I\!I}$ --- Perl, Linux, Array II
 \item $POA_{I\!I}$ --- Perl, OSX, Array II
\end{enumerate}

The following table will display a comparison of the CPU and real times for
the calculation of 19 $s$-values of the joint probabilities $\big(P_{jk}(a,
b)^s\big)$ for the protein family PF06850 by the four configurations
nominated above. It should be stressed that we are here comparing the times
for calculating the $s$-power with the values of the probabilities
themselves previously calculated and kept on a file.

\begin{sidewaystable}[p]
 \begin{center}
  \caption{\small A comparison of calculation times (CPU and real) for 19
  $s$-powers of joint probabilities only of PF06850 protein family, using
  the four configurations $MLA_I$, $POA_{I\!I}$, $PLA_{I\!I}$, $POH$. \label{tab12}}
  \begingroup
  \everymath{\scriptstyle}
  \small
  \begin{tabular}{|c|c|c|c|c|c|c|c|c|}
   \hline
   \multicolumn{1}{|c|}{\multirow{2}{*}{$\mathbf{s}$}} &
   \multicolumn{4}{c|}{$\mathbf{t_{CPU}}$ (sec)} &
   \multicolumn{4}{c|}{$\mathbf{t_R}$ (sec)} \\
   \cline{2-9}
   \multicolumn{1}{|c|}{} & $MLA_I$ & $POA_{I\!I}$ & $PLA_{I\!I}$ & $POH$ &
   $MLA_I$ & $POA_{I\!I}$ & $PLA_{I\!I}$ & $POH$ \\
   \hline
   $0.1$ & $390.432$ & $33.478$ & $\mathbf{41.633}$ & $24.849$ & $206.646$ &
   $88.527$ & $83.139$ & $26.862$ \\
   \hline
   $0.2$ & $382.887$ & $31.711$ & $26.483$ & $25.093$ & $202.282$ & $80.624$
   & $53.384$ & $26.904$ \\
   \hline
   $0.3$ & $401.269$ & $31.726$ & $25.286$ & $24.751$ & $210.791$ & $80.519$
   & $50.689$ & $27.305$ \\
   \hline
   $0.4$ & $416.168$ & $31.448$ & $26.138$ & $26.345$ & $216.993$ & $79.032$
   & $51.409$ & $27.687$ \\
   \hline
   $0.5$ & $427.572$ & $32.860$ & $\mathbf{25.822}$ & $26.726$ & $221.541$
   & $93.048$ & $51.334$ & $27.528$ \\
   \hline
   $0.6$ & $430.604$ & $33.444$ & $27.013$ & $25.021$ & $223.227$ & $102.317$
   & $52.889$ & $25.408$ \\
   \hline
   $0.7$ & $421.904$ & $31.053$ & $25.814$ & $25.011$ & $218.484$ & $79.255$
   & $51.466$ & $26.414$ \\
   \hline
   $0.8$ & $434.888$ & $31.526$ & $26.725$ & $25.183$ & $224.267$ & $80.469$
   & $53.388$ & $25.668$ \\
   \hline
   $0.9$ & $431.948$ & $31.482$ & $26.895$ & $25.409$ & $223.023$ & $80.002$
   & $53.579$ & $25.536$ \\
   \hline
   $1.0$ & $442.933$ & $32.056$ & $25.990$ & $25.096$ & $224.731$ & $80.917$
   & $51.687$ & $25.640$ \\
   \hline
   $2.0$ & $176.212$ & $32.638$ & $27.089$ & $26.012$ & $147.454$ & $80.751$
   & $54.384$ & $26.960$ \\
   \hline
   $3.0$ & $234.100$ & $31.892$ & $25.766$ & $24.498$ & $174.853$ & $85.853$
   & $51.843$ & $24.717$ \\
   \hline
   $4.0$ & $284.184$ & $31.662$ & $26.515$ & $25.251$ & $181.552$ & $91.837$
   & $52.636$ & $25.718$ \\
   \hline
   $5.0$ & $327.740$ & $32.295$ & $27.516$ & $24.925$ & $178.117$ & $87.486$
   & $54.908$ & $25.814$ \\
   \hline
   $6.0$ & $334.800$ & $32.674$ & $28.126$ & $25.440$ & $194.691$ & $86.569$
   & $54.611$ & $25.847$ \\
   \hline
   $7.0$ & $349.064$ & $31.674$ & $23.908$ & $26.389$ & $195.258$ & $86.215$
   & $49.262$ & $27.745$ \\
   \hline
   $8.0$ & $361.304$ & $33.105$ & $26.020$ & $25.106$ & $195.437$ & $116.601$
   & $51.889$ & $26.735$ \\
   \hline
   $9.0$ & $386.217$ & $31.881$ & $26.208$ & $24.783$ & $197.150$ & $81.372$
   & $53.114$ & $25.155$ \\
   \hline
   $10.0$ & $397.276$ & $32.269$ & $26.125$ & $24.979$ & $197.868$ & $87.963$
   & $52.541$ & $26.504$ \\
   \hline \hline
   Total & $7036.502$ & $611.374$ & $515.072$ & $480.867$ & $3834.365$
   & $1649.357$ & $1028.655$ & $500.147$ \\
   \hline \hline
   Total & $7,522,020.640$ & $653,588.806$ & $550,611.968$ & $514,046.813$ &
   $4,098,936.190$ & $1,763,162.630$ & $1,099,632.200$ & $534,157.143$ \\
   (1069) families & $=87.067$ days & $=7.565$ days & $=6.373$ days & $=5.950$ days &
   $=47.441$ days & $=20.407$ days & $=12.727$ days & $=6.188$ days \\
   \hline
  \end{tabular}
  \endgroup
 \end{center}
\end{sidewaystable}

We can check that the times on table \ref{tab12} seem to be generically
ordered as
\begin{equation}
 t_{MLA_I} > t_{POA_{I\!I}} > t_{PLA_{I\!I}} > t_{POH} \label{eq:eq60}
\end{equation}
From this ordering of computing times, we are then able to consider that the
inconvenience of using the ``hash'' structure which has been emphasized on
section 2, was not circumvented by working with a modified array structure
($A_{I\!I}$ instead of $H$), at least for the Mac Pro machine used in these
calculations. We do not also know if this machine has been even used with an
``overload'' of programs from the assumed part-time job of the experimenter
(maybe 99.99\% time job!). Anyhow, the usual Hash structure of Perl computing
system has delayed the calculation of the Entropy Measures and even with the
help of the modified $A_{I\!I}$ structure, it does not succeed at computing
with this structure if operated on a OSX computing system.

On the other hand, the configuration $MLA_I$ could be chosen for parallelizing
the respective adopted code in a work to be done with supercomputer facilities.
If we try to avoid this kind of computational facility, in the belief that the
problem of classifying the distribution of amino acids of a protein database
in terms of Entropy Measures could be treated with less powerful but very
objective ``weapons'', we should try to look for very fast laptop machines
instead, by working with a Linux operating system, a Perl computing system and
a modified array structure. This means that it would be worthwhile the
continuation of the present work with the $PLA_{I\!I}$ configuration. This is
now in progress and will be published elsewhere.

We summarize the conclusions commented above on tables \ref{tab13}--\ref{tab16}
below for the calculation of CPU and Real times of 19 $s$-powers of joint
probabilities $\big(P_{jk}(a,b)\big)^s$ and the corresponding values of
Havrda-Charvat entropy measures. The necessary times for calculating the joint
probabilities themselves has not been taken into consideration. It would be
very useful to make a comparison of the results of table \ref{tab10} with those
on tables \ref{tab14}, \ref{tab16}, and table \ref{tab12} with tables \ref{tab13},
\ref{tab15} as well.

As a last remark of this section, we shall take into consideration,
the restrictions of $s \leq 1$ for working with Jaccard Entropy
measures and we calculate the total CPU and real times for the set
of $s$-values: $s =$ $0.1$, $0.2$, $0.3$, $0.4$, $0.5$, $0.6$, $0.7$,
$0.8$, $0.9$, $1.0$. The results are presented in table \ref{tab17}
below for the $PLA_{I\!I}$ configuration and the calculation of the
Havrda-Charvat entropies.

\begin{sidewaystable}[p]
 \begin{center}
  \caption{\small Calculation of CPU and Real times of 19 $s$-powers of joint
  probabilities $\big(P_{jk}(a,b)\big)^s$ measures for 06 families from 03
  Clans with the $POA_{I\!I}$ configuration. \label{tab13}}
  \begingroup
  \everymath{\scriptstyle}
  \footnotesize
  \begin{tabular}{|c|c|c|c|c|c|c|c|c|c|c|c|c|}
   \hline
   {} & \multicolumn{4}{c|}{CL0028} & \multicolumn{4}{c|}{CL0023}
   & \multicolumn{4}{c|}{CL0257} \\
   \cline{2-13}
   s & \multicolumn{2}{c|}{PF06850} & \multicolumn{2}{c|}{PF00135}
   & \multicolumn{2}{c|}{PF00005} & \multicolumn{2}{c|}{PF13481}
   & \multicolumn{2}{c|}{PF02388} & \multicolumn{2}{c|}{PF09924} \\
   \cline{2-13}
   {} & $t_{CPU}$ (sec) & $t_R$ (sec) & $t_{CPU}$ (sec) & $t_R$ (sec)
   & $t_{CPU}$ (sec) & $t_R$ (sec) & $t_{CPU}$ (sec) & $t_R$ (sec)
   & $t_{CPU}$ (sec) & $t_R$ (sec) & $t_{CPU}$ (sec) & $t_R$ (sec) \\
   \hline
   $0.1$ & $33.478$ & $88.527$ & $46.747$ & $130.014$ & $41.475$ & $169.679$
   & $44.187$ & $171.636$ & $44.716$ & $242.376$ & $46.177$ & $288.259$ \\
   \hline
   $0.2$ & $31.711$ & $80.624$ & $42.157$ & $90.687$ & $36.893$ & $78.893$
   & $38.567$ & $88.754$ & $37.088$ & $88.091$ & $40.709$ & $148.371$ \\
   \hline
   $0.3$ & $31.726$ & $80.519$ & $39.957$ & $82.240$ & $36.929$ & $97.270$
   & $38.800$ & $81.613$ & $37.350$ & $88.664$ & $38.986$ & $119.781$ \\
   \hline
   $0.4$ & $31.448$ & $79.032$ & $41.737$ & $87.934$ & $36.252$ & $80.428$
   & $38.500$ & $78.757$ & $35.592$ & $80.337$ & $36.773$ & $87.217$ \\
   \hline
   $0.5$ & $32.860$ & $93.048$ & $41.130$ & $89.997$ & $38.203$ & $93.595$
   & $37.618$ & $80.179$ & $35.117$ & $78.476$ & $35.534$ & $82.528$ \\
   \hline
   $0.6$ & $33.944$ & $102.317$ & $41.417$ & $120.420$ & $37.143$ & $81.289$
   & $38.387$ & $81.667$ & $45.900$ & $501.222$ & $35.830$ & $83.439$ \\
   \hline
   $0.7$ & $31.053$ & $79.255$ & $40.422$ & $78.531$ & $36.386$ & $84.421$
   & $43.216$ & $562.659$ & $43.452$ & $259.861$ & $35.261$ & $72.605$ \\
   \hline
   $0.8$ & $31.526$ & $80.469$ & $41.556$ & $118.519$ & $36.955$ & $79.543$
   & $41.862$ & $148.028$ & $35.047$ & $78.469$ & $35.718$ & $85.142$ \\
   \hline
   $0.9$ & $31.482$ & $80.002$ & $41.386$ & $81.747$ & $36.811$ & $81.025$
   & $35.518$ & $78.547$ & $35.540$ & $74.570$ & $40.534$ & $204.673$ \\
   \hline
   $1.0$ & $32.056$ & $80.917$ & $40.724$ & $80.958$ & $37.234$ & $80.587$
   & $36.610$ & $87.848$ & $36.353$ & $78.767$ & $39.095$ & $125.292$ \\
   \hline
   $2.0$ & $32.638$ & $80.751$ & $40.701$ & $79.667$ & $38.452$ & $79.336$
   & $36.916$ & $111.199$ & $36.658$ & $81.415$ & $40.415$ & $182.552$ \\
   \hline
   $3.0$ & $31.892$ & $85.853$ & $40.822$ & $79.293$ & $38.223$ & $80.688$
   & $37.356$ & $84.312$ & $36.658$ & $81.415$ & $39.774$ & $157.090$ \\
   \hline
   $4.0$ & $31.662$ & $91.837$ & $41.208$ & $79.825$ & $37.936$ & $98.905$
   & $37.000$ & $86.906$ & $36.012$ & $77.741$ & $38.794$ & $140.857$ \\
   \hline
   $5.0$ & $32.295$ & $87.486$ & $41.059$ & $82.191$ & $38.293$ & $83.289$
   & $36.245$ & $80.873$ & $35.308$ & $77.225$ & $39.927$ & $147.368$ \\
   \hline
   $6.0$ & $32.674$ & $86.569$ & $41.215$ & $86.311$ & $37.601$ & $82.375$
   & $36.395$ & $97.307$ & $35.748$ & $76.573$ & $39.327$ & $154.829$ \\
   \hline
   $7.0$ & $31.674$ & $86.215$ & $41.290$ & $88.714$ & $38.341$ & $80.991$
   & $36.724$ & $95.148$ & $36.194$ & $75.341$ & $39.826$ & $153.204$ \\
   \hline
   $8.0$ & $33.105$ & $116.601$ & $40.984$ & $83.157$ & $37.756$ & $81.073$
   & $40.524$ & $105.466$ & $36.716$ & $82.921$ & $41.056$ & $175.943$ \\
   \hline
   $9.0$ & $31.881$ & $81.372$ & $41.469$ & $113.561$ & $38.748$ & $82.499$
   & $37.874$ & $94.981$ & $40.341$ & $121.311$ & $39.847$ & $144.595$ \\
   \hline
   $10.0$ & $32.269$ & $87.963$ & $41.466$ & $89.366$ & $37.807$ & $81.043$
   & $37.134$ & $88.005$ & $40.053$ & $136.243$ & $40.333$ & $171.395$ \\
   \hline \hline
   Total & $611.374$ & $1649.357$ & $787.447$ & $1743.132$ & $717.438$
   & $1676.929$ & $729.433$ & $2303.885$ & $719.843$ & $2381.018$ & $743.916$
   & $2725.140$ \\
   \hline
  \end{tabular}
  \endgroup
 \end{center}
\end{sidewaystable}

\begin{sidewaystable}[p]
 \begin{center}
  \caption{\small Calculation of CPU and Real times of Havrda-Charvat Entropy
  measures for 06 families from 03 Clans with the $POA_{I\!I}$
  configuration. \label{tab14}}
  \begingroup
  \everymath{\scriptstyle}
  \footnotesize
  \begin{tabular}{|c|c|c|c|c|c|c|c|c|c|c|c|c|}
   \hline
   {} & \multicolumn{4}{c|}{CL0028} & \multicolumn{4}{c|}{CL0023}
   & \multicolumn{4}{c|}{CL0257} \\
   \cline{2-13}
   s & \multicolumn{2}{c|}{PF06850} & \multicolumn{2}{c|}{PF00135}
   & \multicolumn{2}{c|}{PF00005} & \multicolumn{2}{c|}{PF13481}
   & \multicolumn{2}{c|}{PF02388} & \multicolumn{2}{c|}{PF09924} \\
   \cline{2-13}
   {} & $t_{CPU}$ (sec) & $t_R$ (sec) & $t_{CPU}$ (sec) & $t_R$ (sec)
   & $t_{CPU}$ (sec) & $t_R$ (sec) & $t_{CPU}$ (sec) & $t_R$ (sec)
   & $t_{CPU}$ (sec) & $t_R$ (sec) & $t_{CPU}$ (sec) & $t_R$ (sec) \\
   \hline
   $0.1$ & $19.451$ & $26.196$ & $22.620$ & $56.287$ & $25.411$ & $121.508$
   & $18.968$ & $32.706$ & $23.380$ & $80.826$ & $22.502$ & $77.826$ \\
   \hline
   $0.2$ & $19.245$ & $24.901$ & $22.851$ & $69.992$ & $23.362$ & $65.508$
   & $18.810$ & $26.835$ & $23.366$ & $87.098$ & $23.330$ & $60.684$ \\
   \hline
   $0.3$ & $19.582$ & $26.001$ & $23.963$ & $60.960$ & $22.598$ & $72.363$
   & $18.602$ & $26.028$ & $20.012$ & $43.734$ & $21.929$ & $48.947$ \\
   \hline
   $0.4$ & $20.194$ & $31.418$ & $22.211$ & $54.562$ & $23.448$ & $63.369$
   & $19.176$ & $30.482$ & $21.218$ & $67.817$ & $22.182$ & $45.480$ \\
   \hline
   $0.5$ & $20.162$ & $31.653$ & $23.163$ & $67.391$ & $22.153$ & $55.173$
   & $19.281$ & $34.495$ & $21.037$ & $55.237$ & $23.200$ & $62.384$ \\
   \hline
   $0.6$ & $20.941$ & $34.979$ & $23.961$ & $62.158$ & $22.342$ & $57.081$
   & $20.794$ & $44.558$ & $21.087$ & $54.900$ & $22.477$ & $62.421$ \\
   \hline
   $0.7$ & $20.805$ & $34.057$ & $23.679$ & $82.669$ & $19.587$ & $35.750$
   & $19.982$ & $41.314$ & $20.375$ & $32.699$ & $22.398$ & $66.438$ \\
   \hline
   $0.8$ & $20.900$ & $34.146$ & $22.787$ & $60.924$ & $19.081$ & $34.116$
   & $18.890$ & $28.054$ & $20.167$ & $32.685$ & $23.056$ & $61.376$ \\
   \hline
   $0.9$ & $20.909$ & $34.568$ & $22.808$ & $54.890$ & $19.030$ & $33.258$
   & $18.894$ & $29.712$ & $19.422$ & $26.934$ & $23.543$ & $65.099$ \\
   \hline
   $1.0$ & $21.353$ & $35.024$ & $22.860$ & $51.308$ & $19.789$ & $32.218$
   & $19.869$ & $37.922$ & $21.076$ & $35.774$ & $22.528$ & $52.209$ \\
   \hline
   $2.0$ & $20.505$ & $32.920$ & $21.085$ & $46.921$ & $19.672$ & $33.778$
   & $19.083$ & $62.081$ & $22.530$ & $82.336$ & $21.783$ & $51.656$ \\
   \hline
   $3.0$ & $23.923$ & $58.898$ & $22.020$ & $47.298$ & $18.788$ & $31.926$
   & $19.097$ & $35.399$ & $24.210$ & $83.371$ & $21.636$ & $69.905$ \\
   \hline
   $4.0$ & $24.622$ & $68.692$ & $21.954$ & $50.909$ & $18.556$ & $26.512$
   & $19.208$ & $32.014$ & $24.300$ & $112.613$ & $21.458$ & $49.931$ \\
   \hline
   $5.0$ & $24.221$ & $60.107$ & $22.131$ & $58.975$ & $19.424$ & $33.185$
   & $18.231$ & $25.898$ & $24.199$ & $95.807$ & $22.011$ & $55.109$ \\
   \hline
   $6.0$ & $24.475$ & $64.843$ & $22.911$ & $72.004$ & $20.582$ & $38.006$
   & $18.330$ & $25.959$ & $22.741$ & $61.516$ & $21.857$ & $67.976$ \\
   \hline
   $7.0$ & $24.593$ & $70.336$ & $22.779$ & $51.309$ & $20.564$ & $38.390$
   & $19.583$ & $30.719$ & $23.222$ & $88.997$ & $23.018$ & $85.054$ \\
   \hline
   $8.0$ & $25.613$ & $83.423$ & $20.058$ & $34.861$ & $20.460$ & $35.589$
   & $19.977$ & $37.130$ & $23.825$ & $110.933$ & $24.474$ & $90.322$ \\
   \hline
   $9.0$ & $24.139$ & $73.873$ & $21.418$ & $44.709$ & $19.274$ & $32.129$
   & $18.871$ & $29.089$ & $24.207$ & $106.471$ & $23.349$ & $74.724$ \\
   \hline
   $10.0$ & $25.785$ & $101.370$ & $22.183$ & $46.878$ & $23.625$ & $87.442$
   & $22.652$ & $64.895$ & $23.779$ & $86.225$ & $21.802$ & $57.444$ \\
   \hline \hline
   Total & $421.418$ & $927.405$ & $427.442$ & $1075.005$ & $397.746$
   & $927.301$ & $368.298$ & $675.290$ & $424.153$ & $1345.973$ & $428.533$
   & $1204.985$ \\
   \hline
  \end{tabular}
  \endgroup
 \end{center}
\end{sidewaystable}
  
\begin{sidewaystable}[p]
 \begin{center}
  \caption{\small Calculation of CPU and Real times of 19 $s$-powers of joint
  probabilities $\big(P_{jk}(a,b)\big)^s$ measures for 06 families from 03
  Clans with the $PLA_{I\!I}$ configuration. \label{tab15}}
  \begingroup
  \everymath{\scriptstyle}
  \footnotesize
  \begin{tabular}{|c|c|c|c|c|c|c|c|c|c|c|c|c|}
   \hline
   {} & \multicolumn{4}{c|}{CL0028} & \multicolumn{4}{c|}{CL0023}
   & \multicolumn{4}{c|}{CL0257} \\
   \cline{2-13}
   s & \multicolumn{2}{c|}{PF06850} & \multicolumn{2}{c|}{PF00135}
   & \multicolumn{2}{c|}{PF00005} & \multicolumn{2}{c|}{PF13481}
   & \multicolumn{2}{c|}{PF02388} & \multicolumn{2}{c|}{PF09924} \\
   \cline{2-13}
   {} & $t_{CPU}$ (sec) & $t_R$ (sec) & $t_{CPU}$ (sec) & $t_R$ (sec)
   & $t_{CPU}$ (sec) & $t_R$ (sec) & $t_{CPU}$ (sec) & $t_R$ (sec)
   & $t_{CPU}$ (sec) & $t_R$ (sec) & $t_{CPU}$ (sec) & $t_R$ (sec) \\
   \hline
   $0.1$ & $41.633$ & $83.139$ & $43.135$ & $83.901$ & $44.003$ & $84.655$
   & $26.613$ & $52.365$ & $27.452$ & $53.350$ & $43.132$ & $83.249$ \\
   \hline
   $0.2$ & $26.483$ & $53.384$ & $26.373$ & $50.442$ & $26.610$ & $52.904$
   & $26.599$ & $51.822$ & $43.199$ & $83.497$ & $25.762$ & $50.203$ \\
   \hline
   $0.3$ & $25.286$ & $50.689$ & $26.644$ & $50.992$ & $29.965$ & $53.250$
   & $24.946$ & $48.846$ & $25.498$ & $50.303$ & $25.904$ & $50.737$ \\
   \hline
   $0.4$ & $26.138$ & $51.409$ & $25.255$ & $48.576$ & $26.696$ & $52.482$
   & $27.769$ & $53.837$ & $25.753$ & $51.013$ & $25.684$ & $50.371$ \\
   \hline
   $0.5$ & $25.822$ & $51.837$ & $26.041$ & $50.492$ & $26.648$ & $52.171$
   & $25.026$ & $48.927$ & $25.727$ & $49.887$ & $25.513$ & $48.793$ \\
   \hline
   $0.6$ & $27.013$ & $52.889$ & $25.778$ & $50.324$ & $24.912$ & $49.340$
   & $26.062$ & $51.312$ & $26.066$ & $51.417$ & $25.435$ & $49.686$ \\
   \hline
   $0.7$ & $25.814$ & $51.466$ & $25.496$ & $49.954$ & $25.284$ & $50.268$
   & $23.698$ & $48.134$ & $25.723$ & $50.760$ & $25.822$ & $51.122$ \\
   \hline
   $0.8$ & $26.725$ & $53.388$ & $26.254$ & $50.865$ & $25.456$ & $50.870$
   & $25.816$ & $51.196$ & $26.883$ & $52.511$ & $29.837$ & $57.367$ \\
   \hline
   $0.9$ & $26.895$ & $53.579$ & $23.740$ & $47.296$ & $26.237$ & $51.725$
   & $28.441$ & $54.743$ & $25.237$ & $50.532$ & $27.951$ & $54.289$ \\
   \hline
   $1.0$ & $25.990$ & $51.687$ & $27.225$ & $54.384$ & $26.014$ & $51.969$
   & $27.148$ & $52.701$ & $23.672$ & $48.547$ & $26.314$ & $51.932$ \\
   \hline
   $2.0$ & $27.089$ & $54.384$ & $26.237$ & $50.335$ & $27.756$ & $54.761$
   & $26.424$ & $52.503$ & $24.811$ & $49.859$ & $25.846$ & $51.125$ \\
   \hline
   $3.0$ & $25.766$ & $51.843$ & $27.342$ & $52.508$ & $25.135$ & $50.954$
   & $27.753$ & $53.650$ & $25.072$ & $49.142$ & $25.727$ & $50.990$ \\
   \hline
   $4.0$ & $26.515$ & $52.636$ & $25.716$ & $50.531$ & $25.106$ & $50.449$
   & $24.871$ & $50.279$ & $25.674$ & $51.536$ & $26.897$ & $52.227$ \\
   \hline
   $5.0$ & $27.516$ & $54.908$ & $26.002$ & $50.390$ & $24.666$ & $49.463$
   & $23.973$ & $47.554$ & $25.675$ & $51.032$ & $27.810$ & $53.820$ \\
   \hline
   $6.0$ & $28.126$ & $54.611$ & $27.441$ & $53.032$ & $26.014$ & $52.209$
   & $25.676$ & $50.426$ & $26.235$ & $51.375$ & $26.180$ & $51.346$ \\
   \hline
   $7.0$ & $23.908$ & $49.262$ & $26.812$ & $51.556$ & $24.696$ & $49.783$
   & $25.519$ & $50.741$ & $25.863$ & $50.995$ & $25.593$ & $50.611$ \\
   \hline
   $8.0$ & $26.020$ & $51.889$ & $25.431$ & $49.135$ & $26.757$ & $52.518$
   & $26.369$ & $49.668$ & $26.401$ & $52.682$ & $25.084$ & $50.596$ \\
   \hline
   $9.0$ & $26.208$ & $53.114$ & $25.856$ & $50.090$ & $25.803$ & $51.253$
   & $24.628$ & $47.880$ & $25.379$ & $51.014$ & $27.049$ & $52.891$ \\
   \hline
   $10.0$ & $26.125$ & $52.541$ & $24.963$ & $47.402$ & $24.369$ & $48.065$
   & $42.270$ & $83.281$ & $24.878$ & $49.926$ & $24.563$ & $48.616$ \\
   \hline \hline
   Total & $515.072$ & $1028.655$ & $511.741$ & $992.205$ & $509.127$
   & $1009.089$ & $509.601$ & $999.865$ & $505.198$ & $999.378$ & $516.103$
   & $1009.971$ \\
   \hline
  \end{tabular}
  \endgroup
 \end{center}
\end{sidewaystable}

\begin{sidewaystable}[p]
 \begin{center}
  \caption{\small Calculation of CPU and Real times of Havrda-Charvat Entropy
  measures for 06 families from 03 Clans with the $PLA_{I\!I}$
  configuration. \label{tab16}}
  \begingroup
  \everymath{\scriptstyle}
  \footnotesize
  \begin{tabular}{|c|c|c|c|c|c|c|c|c|c|c|c|c|}
   \hline
   {} & \multicolumn{4}{c|}{CL0028} & \multicolumn{4}{c|}{CL0023}
   & \multicolumn{4}{c|}{CL0257} \\
   \cline{2-13}
   s & \multicolumn{2}{c|}{PF06850} & \multicolumn{2}{c|}{PF00135}
   & \multicolumn{2}{c|}{PF00005} & \multicolumn{2}{c|}{PF13481}
   & \multicolumn{2}{c|}{PF02388} & \multicolumn{2}{c|}{PF09924} \\
   \cline{2-13}
   {} & $t_{CPU}$ (sec) & $t_R$ (sec) & $t_{CPU}$ (sec) & $t_R$ (sec)
   & $t_{CPU}$ (sec) & $t_R$ (sec) & $t_{CPU}$ (sec) & $t_R$ (sec)
   & $t_{CPU}$ (sec) & $t_R$ (sec) & $t_{CPU}$ (sec) & $t_R$ (sec) \\
   \hline
   $0.1$ & $21.219$ & $26.895$ & $18.698$ & $23.442$ & $19.383$ & $24.675$
   & $21.351$ & $26.966$ & $19.059$ & $24.560$ & $21.451$ & $26.829$ \\
   \hline
   $0.2$ & $22.604$ & $28.014$ & $21.109$ & $26.536$ & $20.336$ & $25.665$
   & $19.555$ & $24.562$ & $19.370$ & $24.681$ & $22.504$ & $27.855$ \\
   \hline
   $0.3$ & $19.056$ & $24.130$ & $20.692$ & $25.663$ & $21.934$ & $27.466$
   & $21.084$ & $26.236$ & $19.317$ & $24.934$ & $19.681$ & $24.732$ \\
   \hline
   $0.4$ & $21.552$ & $27.442$ & $19.981$ & $25.021$ & $18.794$ & $24.142$
   & $21.065$ & $26.485$ & $20.878$ & $26.291$ & $21.947$ & $26.979$ \\
   \hline
   $0.5$ & $31.952$ & $88.406$ & $21.161$ & $26.374$ & $19.123$ & $24.309$
   & $22.534$ & $27.851$ & $19.729$ & $25.011$ & $21.888$ & $27.168$ \\
   \hline
   $0.6$ & $20.195$ & $25.362$ & $21.006$ & $26.371$ & $21.426$ & $30.303$
   & $20.640$ & $26.050$ & $19.424$ & $24.279$ & $20.750$ & $26.175$ \\
   \hline
   $0.7$ & $19.168$ & $23.924$ & $20.989$ & $26.482$ & $21.440$ & $27.129$
   & $19.370$ & $24.287$ & $19.274$ & $24.779$ & $21.545$ & $26.830$ \\
   \hline
   $0.8$ & $21.982$ & $27.362$ & $19.760$ & $24.681$ & $20.662$ & $25.948$
   & $21.324$ & $26.648$ & $21.685$ & $26.964$ & $21.644$ & $27.160$ \\
   \hline
   $0.9$ & $21.356$ & $27.235$ & $21.251$ & $26.550$ & $19.518$ & $24.880$
   & $21.276$ & $26.736$ & $21.283$ & $26.981$ & $19.217$ & $23.910$ \\
   \hline
   $1.0$ & $20.206$ & $25.608$ & $20.914$ & $26.173$ & $20.726$ & $25.877$
   & $21.430$ & $26.625$ & $20.102$ & $25.198$ & $19.810$ & $25.348$ \\
   \hline
   $2.0$ & $19.435$ & $24.882$ & $20.660$ & $25.675$ & $20.798$ & $26.326$
   & $20.301$ & $25.481$ & $19.588$ & $24.837$ & $20.446$ & $25.755$ \\
   \hline
   $3.0$ & $20.549$ & $25.646$ & $19.988$ & $25.479$ & $20.629$ & $25.516$
   & $19.914$ & $25.230$ & $20.438$ & $25.648$ & $19.393$ & $24.815$ \\
   \hline
   $4.0$ & $19.528$ & $24.693$ & $20.649$ & $25.610$ & $19.428$ & $24.637$
   & $20.505$ & $26.001$ & $20.868$ & $26.009$ & $20.060$ & $25.643$ \\
   \hline
   $5.0$ & $19.698$ & $24.824$ & $21.076$ & $26.468$ & $19.865$ & $24.753$
   & $19.120$ & $24.267$ & $19.466$ & $24.423$ & $21.760$ & $27.244$ \\
   \hline
   $6.0$ & $20.809$ & $26.319$ & $20.352$ & $25.914$ & $19.507$ & $24.362$
   & $20.110$ & $25.587$ & $20.708$ & $25.961$ & $21.240$ & $26.665$ \\
   \hline
   $7.0$ & $20.287$ & $25.951$ & $21.073$ & $26.459$ & $19.482$ & $24.861$
   & $18.272$ & $23.535$ & $19.015$ & $24.460$ & $20.646$ & $25.964$ \\
   \hline
   $8.0$ & $21.427$ & $26.885$ & $19.494$ & $24.421$ & $20.107$ & $25.228$
   & $20.645$ & $26.095$ & $21.039$ & $26.222$ & $20.014$ & $25.155$ \\
   \hline
   $9.0$ & $21.623$ & $27.335$ & $21.554$ & $26.517$ & $19.239$ & $24.529$
   & $20.629$ & $25.929$ & $21.035$ & $26.450$ & $21.086$ & $26.526$ \\
   \hline
   $10.0$ & $20.815$ & $26.379$ & $21.127$ & $26.630$ & $19.927$ & $25.340$
   & $19.781$ & $25.063$ & $21.438$ & $27.019$ & $17.286$ & $22.390$ \\
   \hline \hline
   Total & $403.461$ & $557.294$ & $391.524$ & $490.466$ & $382.324$
   & $485.946$ & $388.906$ & $489.634$ & $383.716$ & $484.707$ & $392.368$
   & $493.143$ \\
   \hline
  \end{tabular}
  \endgroup
 \end{center}
\end{sidewaystable}

\begin{sidewaystable}[p]
 \begin{center}
  \caption{\small Calculation of CPU and Real times of Havrda-Charvat
  Entropy measures for 06 families from 03 Clans with parameters
  $0 < s \leq 1$ and the $PLA_{I\!I}$ configuration. \label{tab17}}
  \begingroup
  \everymath{\scriptstyle}
  \footnotesize
  \begin{tabular}{|c|c|c|c|c|c|c|c|c|c|c|c|c|}
   \hline
   {} & \multicolumn{4}{c|}{CL0028} & \multicolumn{4}{c|}{CL0023}
   & \multicolumn{4}{c|}{CL0257} \\
   \cline{2-13}
   s & \multicolumn{2}{c|}{PF06850} & \multicolumn{2}{c|}{PF00135}
   & \multicolumn{2}{c|}{PF00005} & \multicolumn{2}{c|}{PF13481}
   & \multicolumn{2}{c|}{PF02388} & \multicolumn{2}{c|}{PF09924} \\
   \cline{2-13}
   {} & $t_{CPU}$ (sec) & $t_R$ (sec) & $t_{CPU}$ (sec) & $t_R$ (sec)
   & $t_{CPU}$ (sec) & $t_R$ (sec) & $t_{CPU}$ (sec) & $t_R$ (sec)
   & $t_{CPU}$ (sec) & $t_R$ (sec) & $t_{CPU}$ (sec) & $t_R$ (sec) \\
   \hline
   $0.1$ & $21.219$ & $26.895$ & $18.698$ & $23.442$ & $19.383$ & $24.675$
   & $21.351$ & $26.966$ & $19.059$ & $24.560$ & $21.451$ & $26.829$ \\
   \hline
   $0.2$ & $22.604$ & $28.014$ & $21.109$ & $26.536$ & $20.336$ & $25.665$
   & $19.555$ & $24.562$ & $19.370$ & $24.681$ & $22.504$ & $27.855$ \\
   \hline
   $0.3$ & $19.056$ & $24.130$ & $20.692$ & $25.663$ & $21.934$ & $27.466$
   & $21.084$ & $26.236$ & $19.317$ & $24.934$ & $19.681$ & $24.732$ \\
   \hline
   $0.4$ & $21.552$ & $27.442$ & $19.981$ & $25.021$ & $18.794$ & $24.142$
   & $21.065$ & $26.485$ & $20.878$ & $26.291$ & $21.947$ & $26.979$ \\
   \hline
   $0.5$ & $31.952$ & $88.406$ & $21.161$ & $26.374$ & $19.123$ & $24.309$
   & $22.534$ & $27.851$ & $19.729$ & $25.011$ & $21.888$ & $27.168$ \\
   \hline
   $0.6$ & $20.195$ & $25.362$ & $21.006$ & $26.371$ & $21.426$ & $30.303$
   & $20.640$ & $26.050$ & $19.424$ & $24.279$ & $20.750$ & $26.175$ \\
   \hline
   $0.7$ & $19.168$ & $23.924$ & $20.989$ & $26.482$ & $21.440$ & $27.129$
   & $19.370$ & $24.287$ & $19.274$ & $24.779$ & $21.545$ & $26.830$ \\
   \hline
   $0.8$ & $21.982$ & $27.362$ & $19.760$ & $24.681$ & $20.662$ & $25.948$
   & $21.324$ & $26.648$ & $21.685$ & $26.964$ & $21.644$ & $27.160$ \\
   \hline
   $0.9$ & $21.356$ & $27.235$ & $21.251$ & $26.550$ & $19.518$ & $24.880$
   & $21.276$ & $26.736$ & $21.283$ & $26.981$ & $19.217$ & $23.910$ \\
   \hline
   $1.0$ & $20.206$ & $25.608$ & $20.914$ & $26.173$ & $20.726$ & $25.877$
   & $21.430$ & $26.625$ & $20.102$ & $25.198$ & $19.810$ & $25.348$ \\
   \hline \hline
   Total & $219.290$ & $324.380$ & $205.561$ & $257.293$ & $203.342$
   & $260.394$ & $209.629$ & $262.446$ & $200.121$ & $253.678$ & $210.437$
   & $262.986$ \\
   \hline \hline
   Grand & \multicolumn{1}{c|}{\multirow{2}{*}{$1284.343$}}
   & \multicolumn{1}{c|}{\multirow{2}{*}{$1895.873$}}
   & \multicolumn{1}{c|}{\multirow{2}{*}{$1199.098$}}
   & \multicolumn{1}{c|}{\multirow{2}{*}{$1727.300$}}
   & \multicolumn{1}{c|}{\multirow{2}{*}{$1131.878$}}
   & \multicolumn{1}{c|}{\multirow{2}{*}{$1693.572$}}
   & \multicolumn{1}{c|}{\multirow{2}{*}{$1066.375$}}
   & \multicolumn{1}{c|}{\multirow{2}{*}{$1511.927$}}
   & \multicolumn{1}{c|}{\multirow{2}{*}{$1124.465$}}
   & \multicolumn{1}{c|}{\multirow{2}{*}{$1681.087$}}
   & \multicolumn{1}{c|}{\multirow{2}{*}{$1211.808$}}
   & \multicolumn{1}{c|}{\multirow{2}{*}{$1772.106$}} \\
   Total & \multicolumn{1}{c|}{} & \multicolumn{1}{c|}{}
   & \multicolumn{1}{c|}{} & \multicolumn{1}{c|}{} & \multicolumn{1}{c|}{}
   & \multicolumn{1}{c|}{} & \multicolumn{1}{c|}{} & \multicolumn{1}{c|}{}
   & \multicolumn{1}{c|}{} & \multicolumn{1}{c|}{} & \multicolumn{1}{c|}{}
   & \multicolumn{1}{c|}{} \\
   \hline
  \end{tabular}
  \endgroup
 \end{center}
\end{sidewaystable}

\newpage

The corresponding times for calculating the joint probabilities and
$s$-powers of these have been added up to report the results for the
Havrda-Charvat entropies of the Grand total row. We think that these
times are very well affordable indeed.

\begin{table}[H]
 \begin{center}
  \caption{\small Total CPU and real times for calculating the Entropy measure
  content of a family PF06850 and approximations for Grand
  Total of all sample space. \label{tab18}}
  \begin{tabular}{|c|c|c|}
   \hline
   \multicolumn{1}{|c|}{\multirow{2}{*}{$POA_{I\!I}$}} & Entropy Measures &
   Entropy Measures \\
   {} & $H\!_j(s)$ --- 19 $s$-values & $H\!_{jk}(s)$ ---
   19 $s$-values \\
   \hline
   Total CPU time & $0.358+2.562+0.292$
   & $550.129+611.374$ \\
   (family PF06850) & $=3.212$ sec & $+421.418=1,582.921$ sec \\
   \hline
   Grand Total CPU time & $3,433.628$ sec
   & $1,692,142.549$ sec \\
   (1069 families) & $=0.954$ hs & $=19.585$ days \\
   \hline
   Total Real time & $1.062+9.300+0.332$
   & $593.848+1,649.357+$ \\
   (family PF06850) & $=10.694$ sec & $927.405=3.170.61$ sec \\
   \hline
   Grand Total Real time & $11,431.886$ sec
   & $3,389,382.090$ sec \\
   (1069 families) & $=3.175$ hs & $=39.229$ days \\
   \hline
  \end{tabular}
 \end{center}
\end{table}

\begin{table}[H]
 \begin{center}
  \caption{\small Total CPU and real times for calculating the Entropy measure
  content of a family PF06850 and approximations for Grand
  Total of all sample space. \label{tab19}}
  \begin{tabular}{|c|c|c|}
   \hline
   \multicolumn{1}{|c|}{\multirow{2}{*}{$PLA_{I\!I}$}} & Entropy Measures &
   Entropy Measures \\
   {} & $H\!_j(s)$ --- 19 $s$-values & $H\!_{jk}(s)$ ---
   19 $s$-values \\
   \hline
   Total CPU time & $0.291+1.801+0.640$
   & $787.254+515.072$ \\
   (family PF06850) & $=2.732$ sec & $+403.461=1,705.787$ sec \\
   \hline
   Grand Total CPU time & $2,920.508$ sec
   & $1,823,486.303$ sec \\
   (1069 families) & $=0.811$ hs & $=21.105$ days \\
   \hline
   Total Real time & $0.642+7.261+1.291$
   & $1,068.026+1,028.655$ \\
   (family PF06850) & $=9.194$ sec & $+557.294=2,603.975$ sec \\
   \hline
   Grand Total Real time & $9,828.386$ sec
   & $2,783,649.275$ sec \\
   (1069 families) & $=2.730$ hs & $=32.218$ days \\
   \hline
  \end{tabular}
 \end{center}
\end{table}

\section{Concluding Remarks and Suggestions for Future Work}
The treatment of the distributions of probability of occur in protein
databases is a twofold procedure. We intend to find a way of characterizing
the protein database by values of Entropy Measures in order to provide a
sound discussion to be centered on the maximization of a convenient average
Entropy Measure to represent the entire protein database. We also intend to
derive a partition function in order to derive a thermodynamical theory
associated to the temporal evolution of the database. If the corresponding
evolution of the protein families is assumed to be registered on the
subsequent versions of the database (Table \ref{tab17}), we will then be
able to describe the sought thermodynamical evolution from this theory as
well as to obtain from it the convenient description of all intermediate
Levinthal's stages which seem to be necessary for describing the
folding/unfolding dynamical process.

We summarize this approach by the need of starting from a thermodynamical
theory of the evolution of protein databases via Entropy measures to the
construction of a successful dynamical theory of protein families. In other
words, from the thermodynamics of evolution of a protein database, we will
derive a statistical mechanics to give us physical insight on the
construction of a successful dynamics of protein families.

\end{document}